\documentclass[a4paper,11pt]{article}
\usepackage{pos}
\usepackage{amsmath}
\usepackage{bm}
\usepackage{feynmp-auto}
\usepackage{subfig}
\usepackage[makeroom]{cancel}

\title{\boldmath$B\to D^{(\ast)}\ell\nu$ semileptonic decays at non-zero recoil}

\ShortTitle{\boldmath$B\to D^{(\ast)}\ell\nu$ at non-zero recoil}

\author*[a]{Alejandro Vaquero Avil\'es-Casco}
\affiliation[a]{Department of Physics and Astronomy, University of Utah, \\
                Salt Lake City, UT 84112-0830, USA}
\emailAdd{alexv@unizar.es}

\abstract{$B$ anomalies play a prominent role in Beyond the Standard Model (BSM) physics searches. In particular, the long standing tension between the inclusive and the exclusive
          determinations of the CKM matrix element $|V_{cb}|$ and the current tensions in the $R(D)$--$R(D^\ast)$ plane between theory and experiment have brought the
          $B\to D^{(\ast)}\ell\nu$ semileptonic processes to the spotlight.
          Existing lattice-QCD calculations of the $B\to D\ell\nu$ form factors at non-zero recoil are being complemented with very recent developments in the $B\to D^\ast\ell\nu$
          channel. In this review I discuss recent progress in lattice calculations of $B\to D^{(\ast)}\ell\nu$, as well as the implications of these results for high precision
          determinations of $|V_{cb}|$ and the Lepton Flavor Universality (LFU) ratios $R(D^{(\ast)})$.}

\FullConference{%
  The 39th International Symposium on Lattice Field Theory (Lattice2022),\\
  8-13 August, 2022 \\
  Bonn, Germany
}

\begin{document}
\maketitle

\section{Introduction}
\subsection{The scales of the Standard Model}
The Standard Model (SM) is an extremely successful theory, in the sense that it can predict physical phenomena accurately and precisely.
Nonetheless, there are several indisputable signals that there is life beyond the SM.
The current point of view regards the SM as an Effective Field Theory (EFT), and as such it must have a defining scale,
beyond which the theory breaks down and cannot describe Nature any longer.
The defining scale of the SM is the Electroweak (EW) scale $\approx 10^{11}$ GeV, close to the weak bosons masses or the Higgs mass.
\begin{figure}[h]                                                                                                                                                                  
  \includegraphics[width=\textwidth,angle=0]{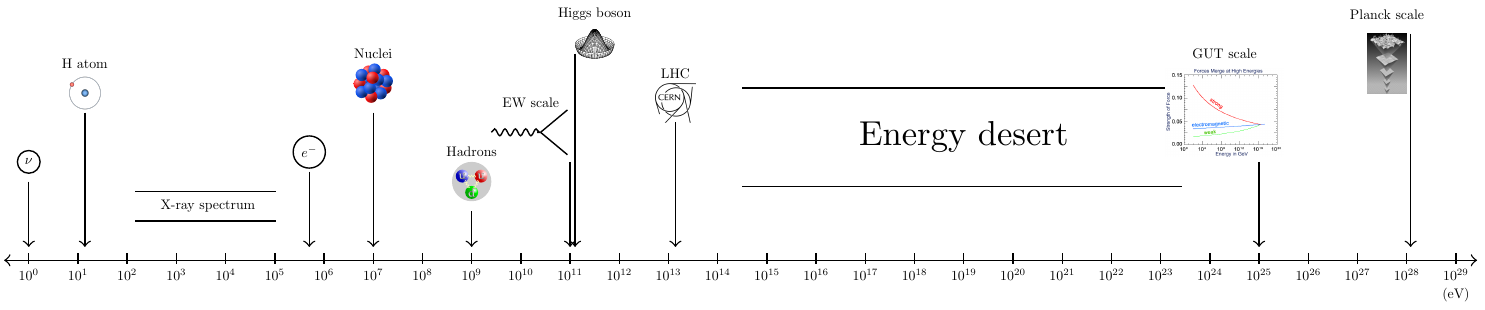}
  \caption{Physical scales of different phenomena.}
  \label{eScales}
\end{figure}
Given this value, one would expect to find New Physics (NP) in high-energy particle accelerators, but as of today not even the mighty LHC has produced a clear signal of NP.
The immediate conclusion is that the true scale where the SM breaks down must be much higher than expected.
Theoretically we do not have many options left: one candidate is the Grand Unification Theory (GUT) scale at $\approx 10^{25}$ eV;
another one would be the Planck mass $\approx 10^{28}$ eV, which lies even further.
This leaves us with a huge \emph{energy desert} (see Fig.~\ref{eScales}), spaning around 15 orders of magnitude, that must be explored in order to find NP.
Accessing such large energies directly is unfeasible with current technology, and we can only hope that the NP scale is close enough to our capabilities.
Alternatively, one can resort to indirect searches: certain observables are calculated at very high precision within the SM framework, and the results are contrasted with experiment.
Since quantum-loop corrections must include contributions from all existing particles, small deviations can indicate the presence of NP.
This approach has the obvious benefit of being extremely sensitive to very high-energy phenomena, but it does not give any information about the nature of the NP.
New models must be constrained taking into account all the existing anomalies, while keeping the rest of the physics untouched.
The task is certainly daunting, and any new information is welcome.

\subsection{The flavor sector of the SM}
\begin{figure}[h]                                                                                                                                                                  
  \includegraphics[width=\textwidth,angle=0]{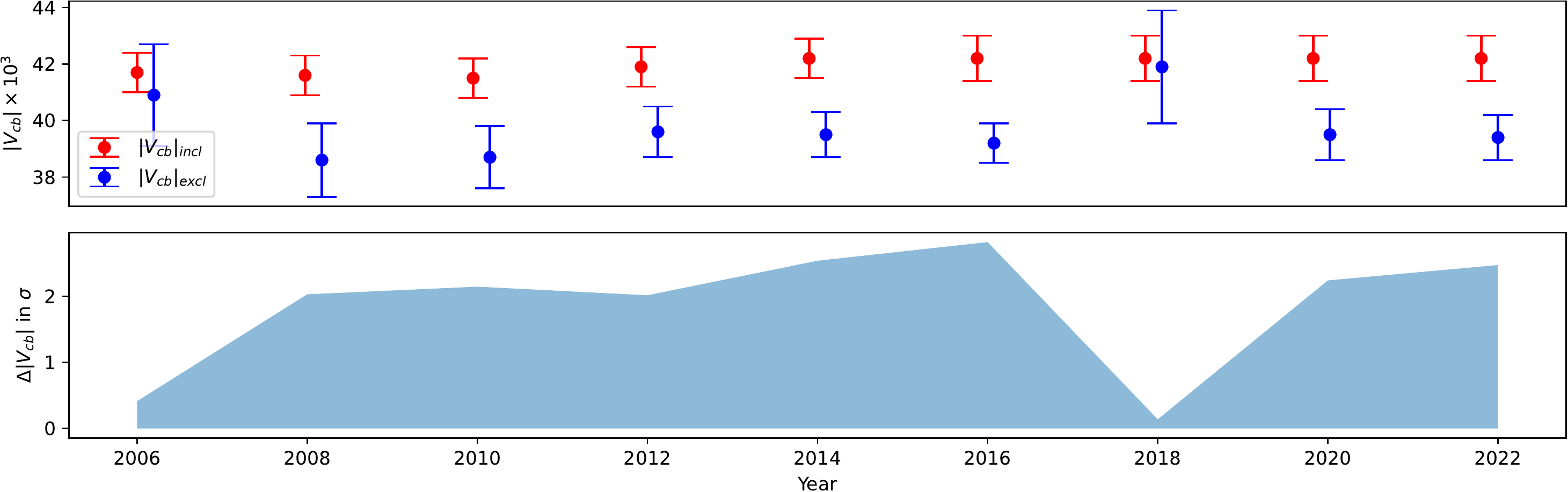}
  \caption{Difference in $\sigma$s between the inclusive and the exclusive determinations of $|V_{cb}|$ plotted against the year.
           All values of $|V_{cb}|$ come from the PDG~\cite{ParticleDataGroup:2006fqo,ParticleDataGroup:2008zun,ParticleDataGroup:2010dbb,ParticleDataGroup:2012pjm,
           ParticleDataGroup:2014cgo,ParticleDataGroup:2016lqr,ParticleDataGroup:2018ovx,ParticleDataGroup:2020ssz,ParticleDataGroup:2022pth}}
  \label{VcbDiff}
\end{figure}

A good place to look for NP is the flavor sector of the SM.
The theory of the weak interactions has many particularities that are missing in both QCD and QED, and some of these make it quite amenable to incorporate NP extensions.
Among them, the existence of a mixing matrix, the CKM matrix, is a prominent one. The CKM matrix relates the flavor eigenstates of the quarks with their mass eigenstates.
Its elements are inputs of the SM, and must be determined by combining experimental results on decays where a particular matrix element is involved, with the theoretical
expressions describing those decays.
One of the most controversial elements of the CKM matrix is $V_{cb}$, due to a long-standing discrepancy between the inclusive and the exclusive determinations.
The discrepancy is not so large that it can be taken as an unmistakable sign of NP, but it is large enough to raise eyebrowns.
We have not been able to clarify the issue for the last 15 years. A summary is gathered in Fig.~\ref{VcbDiff}.

\begin{figure}[h]
  \begin{center}                                                                                                                                                                  
  \includegraphics[width=0.55\textwidth,angle=0]{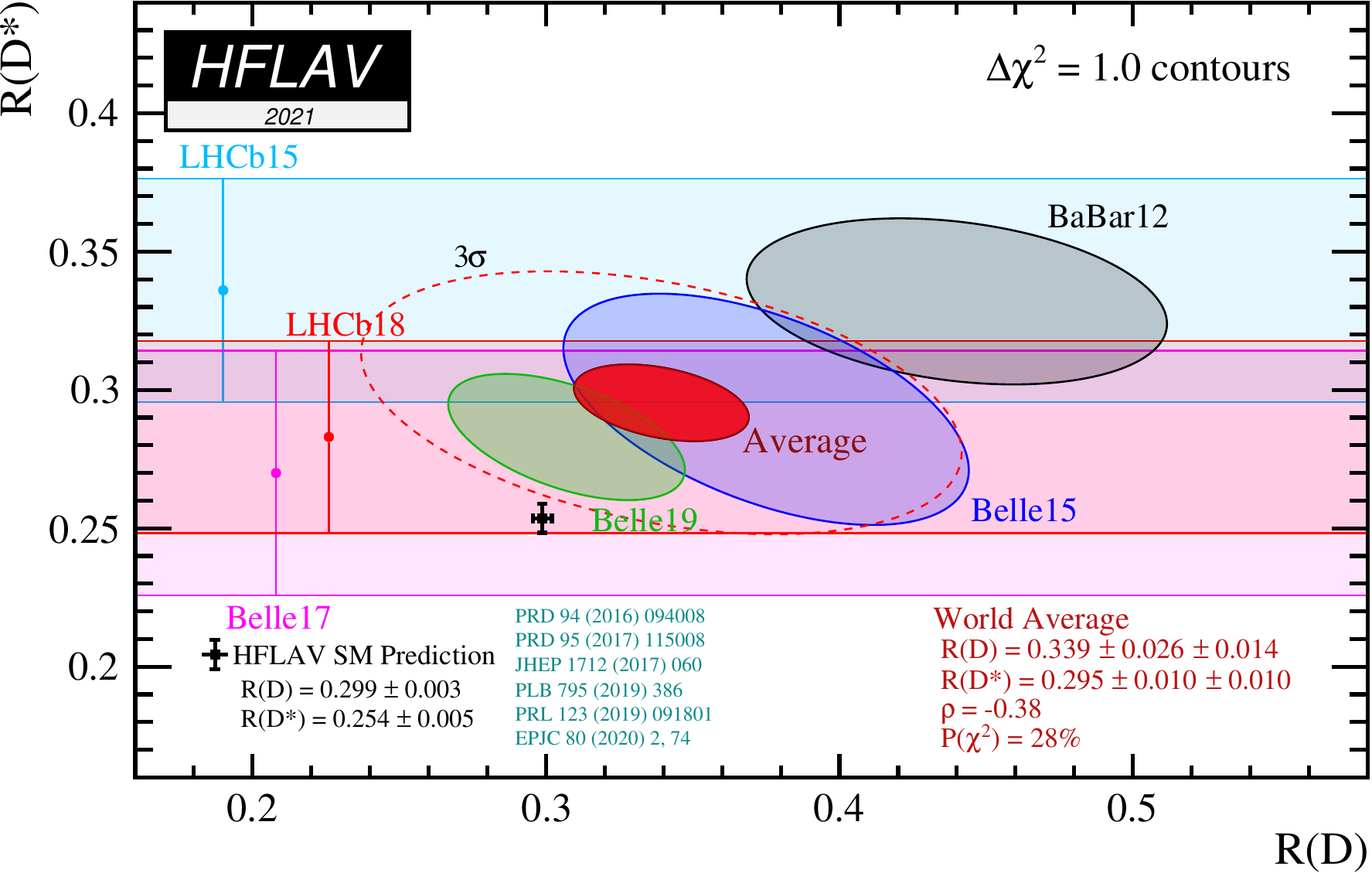}
  \caption{Tensions between theory and experiment in the $R(D)$ vs $R(D^\ast)$ plane.}
  \label{RDvsRDst}
  \end{center}                                                                                                                                                                  
\end{figure}

Another interesting observable that can certainly herald the discovery of NP is the LFU ratio $R(D^{(\ast)})$, defined as
\begin{equation}
R(D^{(\ast)}) = \frac{\mathcal{B}(B\to D^{(\ast)}\tau\nu)}{\mathcal{B}(B\to D^{(\ast)}\ell\nu)},\quad\ell = e,\mu.
\label{RDst}
\end{equation}
These ratios are good candidates for NP searches because many hadronic uncertainties are canceled, hence we can make more precise predictions to compare with experiment.
Currently, the tensions between theory and experiment in the $R(D)$-$R(D^\ast)$ plane are $\approx 3.3\sigma$ (see Fig.~\ref{RDvsRDst}).
An increase in the current precision of experiments is required, especially when comparing the latest Belle measurements~\cite{Belle:2019rba} with the SM theoretical expectations.
On the theory side we would benefit from first principles calculations of $R(D^{(\ast)})$ that could reduce uncontrolled uncertainties. Lattice QCD can provide such calculations.

\subsection{General theory of the $B\to D^{(\ast)}\ell\nu$ decays}\label{SecCT}
The differential decay rate can be written as
\begin{equation}
\underbrace{\frac{d\Gamma}{dw}(B\to D^{(\ast)}\ell\nu)}_{\textrm{Experiment}} = \left[\underbrace{K^{D^{(\ast)}}_1(w)}_{\textrm{Known factors}} \times \underbrace{\left|F(w)\right|^2}_{\textrm{Theory}} 
+ \underbrace{K^{D^{(\ast)}}_2(w,m_\ell)}_{\textrm{Known factors}} \times \underbrace{\left|F_2(w)\right|^2}_{\textrm{Theory}}\right] \times \left|V_{cb}\right|^2,
\label{BtoDstDR}
\end{equation}
where $w=v_{D^{(\ast)}}\cdot v_B$, the product of the four-velocities of the $D^{(\ast)}$ and the $B$ meson is the recoil parameter, which equals $E_{D^{(\ast)}}/M_{D^{(\ast)}}$,
the ratio between the energy and the mass of the $D^{(\ast)}$ meson in the lab frame (keeping the $B$ meson at rest), $K^{D^{(\ast)}}_{1,2}$ comprise
known factors (kinematic factors, constants, etc) defined in the theory, and $F=\mathcal{G},\mathcal{F}$ is the decay amplitude for the $B\to D^{(\ast)}\ell\nu$ decay that must
be calculated from the theory. The term $K^{D^{(\ast)}}|F_2|$ is proportional to the lepton mass $m_\ell$, and it contributes noticeably to the total differential decay rate only
when $m_\ell$ is not negligible with respect to the available $q^2$.
In these two channels this situation happens just for the $\tau$ lepton, and the extra information provided is relevant for the calculation of the LFU ratios.
Exclusive $|V_{cb}|$ determinations are done in the electron and muon channels. Since experiments can only access the lhs of Eq.~\eqref{BtoDstDR}, there is no way to disentangle
the different factors on the rhs of the equation, and theory input is required.
On the other hand, theory gives information on the expression in Eq.~\eqref{BtoDstDR} and $F$, but $|V_{cb}|$ is an input and can't be calculated.
Hence, a combination of theory and experiment is needed to extract $|V_{cb}|$.
 
Experimentally, the $B\to D\ell\nu$ decay suffers from a large background coming from the $D^\ast$, whereas the $B\to D^\ast\ell\nu$ decays enjoy a much better signal to noise ratio.
On top of that, the $B\to D^\ast\ell\nu$ differential decay rate is suppressed at small recoil, since $K_{D^\ast}\propto (w^2-1)^{1/2}$, but the competing channel $B\to D\ell\nu$
is suppressed by a $(w^2-1)^{3/2}$ factor, which is much worse.
In general, the experimental data of these channels is much more precise at large recoil than at low recoil, but it is the low recoil region that matters for $|V_{cb}|$ extraction.

The theoretical situation is reversed: regarding the most precise region where data exists, the low recoil region is more accessible since it involves correlators at
lower momenta, and the errors are more under control.
Moreover, it is much easier to handle the decay to a pseudoscalar on the lattice because it involves just two form factors, as opposed to the four form factors required for
the pseudoscalar dacaying to a vector case.
For that reason, precise calculations of the $B\to D\ell\nu$ form factors in the whole recoil range exist since many years ago, but $B\to D^\ast\ell\nu$ form factors
where restricted to $w=1$ until very recently~\cite{FermilabLattice:2021cdg}\footnote{The first calculation of the $B\to D^\ast\ell\nu$ form factors in the whole recoil range was
published in~\cite{deDivitiis:2008df} as early as 2009 using the quenched approximation.}. The current aim in $B\to D\ell\nu$ calculations is to increase precision and calculate form
factors impacting BSM models, whereas in the $B\to D^\ast\ell\nu$ case calculations of the SM form factors covering the whole kinematic range are starting to emerge.

Both $|V_{cb}|$ and $R(D^{(\ast)})$ are observables at the forefront of BSM physics searches in the intensity frontier,
and large amount of resources are being dedicated to reduce the errors.
$B$-factories like Belle II and the High-Luminosity LHC promise to aggressively decrease uncertainties in the coming years, but all these improvements are useless if
there are no matching efforts coming from theory. This short review aims at giving a clear picture of the status of the calculations of the form factors of the $B\to D^{(\ast)}\ell\nu$
semileptonic decays.

\section{The $B\to D^{(\ast)}\ell\nu$ decay in LQCD}
\subsection{Form factors}
Lattice QCD allows to calculate non-perturbatively all the form factors contributing to the $B\to D^{(\ast)}\ell\nu$ decays.
The form factors are extracted from the relevant correlators,
\begin{align}
\frac{\left\langle D^\ast(p_D)\right|\mathcal{V}^\mu\left|\bar{B}(p_B)\right\rangle}{\sqrt{M_B\,M_D}} =&\quad
\left(v_B^\mu + v_D^\mu\right) \bm{h_+}(w) + \left(v_B^\mu - v_D^\mu\right) \bm{h_-}(w), \\
\frac{\left\langle D^\ast(p_{D^\ast},\epsilon^\nu)\right|\mathcal{V}^\mu\left|\bar{B}(p_B)\right\rangle}{2\sqrt{M_B\,M_{D^\ast}}} =&\quad
\frac{1}{2}\epsilon^{\nu *}\varepsilon^{\mu\nu}_{\,\,\rho\sigma} v_B^\rho v_{D^\ast}^\sigma \bm{h_V}(w), \\
\frac{\left\langle D^\ast(p_{D^\ast},\epsilon^\nu)\right|\mathcal{A}^\mu\left|\bar{B}(p_B)\right\rangle}{2\sqrt{m_B\,m_{D^\ast}}} =&\quad
\frac{i}{2}\epsilon^{\nu *}\left[g^{\mu\nu}\left(1+w\right)\bm{h_{A_1}}(w) - v_B^\nu\left(v_B^\mu \bm{h_{A_2}}(w) + v_{D^\ast}^\mu \bm{h_{A_3}}(w)\right)\right],
\end{align}
where $\mathcal{V}^\mu = $ and $\mathcal{A}^\mu = $ are the vector and axial currents in the continuum, $M_{B,D^{(\ast)}}$ are the rest masses of the relevant mesons and
$p_{B,D^{(\ast)}}$ their momenta, $\varepsilon^{\mu\nu}_{\,\,\rho\sigma}$ is the fully antisymmetric tensor, $\epsilon^{\nu *}$ represents the polarization vector in the case of the
$D^\ast$, and the $h_X(w)$, which have been highlighted in bold, are the different form factors, to be calculated in the lattice.

Complete knowledge of the $h_X(w)$ form factors enables us to calculate the decay amplitudes $\mathcal{G}(w)$ and $\mathcal{F}(w)$ ($F(w)$ in Eq.~\eqref{BtoDstDR}).
This, combined with experimental input are the necessary ingredients to perform an exclusive determination of $|V_{cb}|$.

BSM physics calculations require extra form factors that can also be extracted from the lattice. Collaborations have started to look into these since very recently. As a result,
there is not much data available yet, and most of them is preliminary. Most efforts are going to compute the tensor form factor in the $B\to D\ell\nu$ decay, which is better understood
in LQCD,
\begin{equation}
\frac{\left\langle D^\ast(p_D)\right|\mathcal{T}^{\mu\nu}\left|\bar{B}(p_B)\right\rangle}{\sqrt{M_B\,M_D}} = i\left(v_D^\mu v_B^\nu - v_D^\nu v_B^\mu\right) \bm{h_T}(w).
\end{equation} 

The LFU ratios are different in the sense that no experimental input is required to calculate them from theory.
Integrating Eqs.\eqref{BtoDstDR} in the whole recoil range to obtain the total branching fraction, and explicitly writing the ratio, we get
\begin{equation}
R(D^{(\ast)}) = \frac{\int_1^{w_{\textrm{Max},\tau}} dw\, \left[K_1^{D^{(\ast)}}(w)\left|F(w)\right|^2 + K_2^{D^{(\ast)}}(w,m_\tau)\left|F_2(w)\right|^2\right] \times \xcancel{\left|V_{cb}\right|^2}}
                     {\int_1^{w_{\textrm{Max}}} dw\,K_1^{D^{(\ast)}}(w)\left|F(w)\right|^2\times \xcancel{\left|V_{cb}\right|^2}}
\end{equation}
where $|V_{cb}|$ cancels out.
The possibility of comparing an observable with low hadronic uncertainties, calculated from first principles in LQCD, with experiment is extremely attractive.

\subsection{Parametrizations}

As explained in Sec.~\ref{SecCT}, the exclusive determination of $|V_{cb}|$ requires a combination of experimental data, precise at large recoil, with theoretical data, precise at
small recoil. The usage of parametrizations for the form factors bridges this gap.
Parametrizations also provide an ansatz based on theoretical considerations for the fits to form factor data, and impose bounds on the shapes of the different form factors.

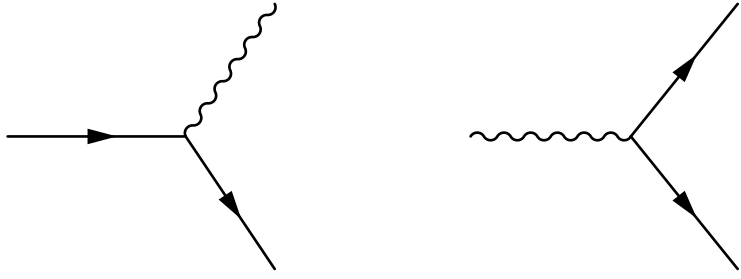
\begin{figure}
  \centering
  \subfloat{
    \begin{fmffile}{SL}
    \begin{fmfgraph*}(100,100)
      \fmfstraight
      \fmfleft{i0}
      \fmfright{o0,o1}
      \fmf{fermion,label=$B$,         label.dist=10,tension=1.5}       {i0,v0}
      \fmf{fermion,label=$D^{(\ast)}$,label.dist=10,tension=1.5}       {v0,o0}
      %\fmffreeze
      \fmf{photon,tension=1.5,label=$W^\pm$,label.side=left,label.dist=5}{v0,o1}
    \end{fmfgraph*}
    \end{fmffile}
  }\hspace{2cm}
  \subfloat{
    \begin{fmffile}{PP}
    \begin{fmfgraph*}(100,100)
      \fmfstraight
      \fmfleft{i2}
      \fmfright{o2,o3}
      \fmf{photon,tension=2,label=$W^\pm$,label.side=left,label.dist=5}{i2,v1}
      \fmf{fermion,label=$B$,         label.dist=10,tension=1.5}       {v1,o2}
      \fmf{fermion,label=$D^{(\ast)}$,label.dist=10,tension=1.5}       {v1,o3}
      %\fmffreeze
    \end{fmfgraph*}
    \end{fmffile}
  }
  \caption{{\bf Left:} semileptonic process. {\bf Right:} pair-production process.\label{SL-PP}}
\end{figure}
Let's consider the processes in Fig.~\ref{SL-PP}. One can connect the semileptonic decay to the pair-production process by a simple rotation.
In the language of form factors this is translated as an analytical continuation to a different region of $q^2$: the semileptonic decay is possible for $q^2\in [m_\ell^2, t_-]$,
with $t_\pm = (m_B\pm m_{D^{(\ast)}})^2$, whereas the pair-production process is only possible if $q^2 \in [t_+, \infty)$.
The semileptonic region can only be accessed with non-perturbative techniques, but the pair-production region is amenable to theoretical calculations using dispersion relations.
One can establish bounds on the form factors by taking into account the different channels that contribute to the vacuum polarization tensor, calculated from the correlator
of two vector or two axial currents. These bounds calculated in the pair-production region are then translated into bounds in the semileptonic region via analytical continuation.
This is the basis of most parametrizations for these decays.

\begin{figure}[h]
  \begin{center}                                                                                                                                                                  
  \includegraphics[width=0.5\textwidth,angle=0]{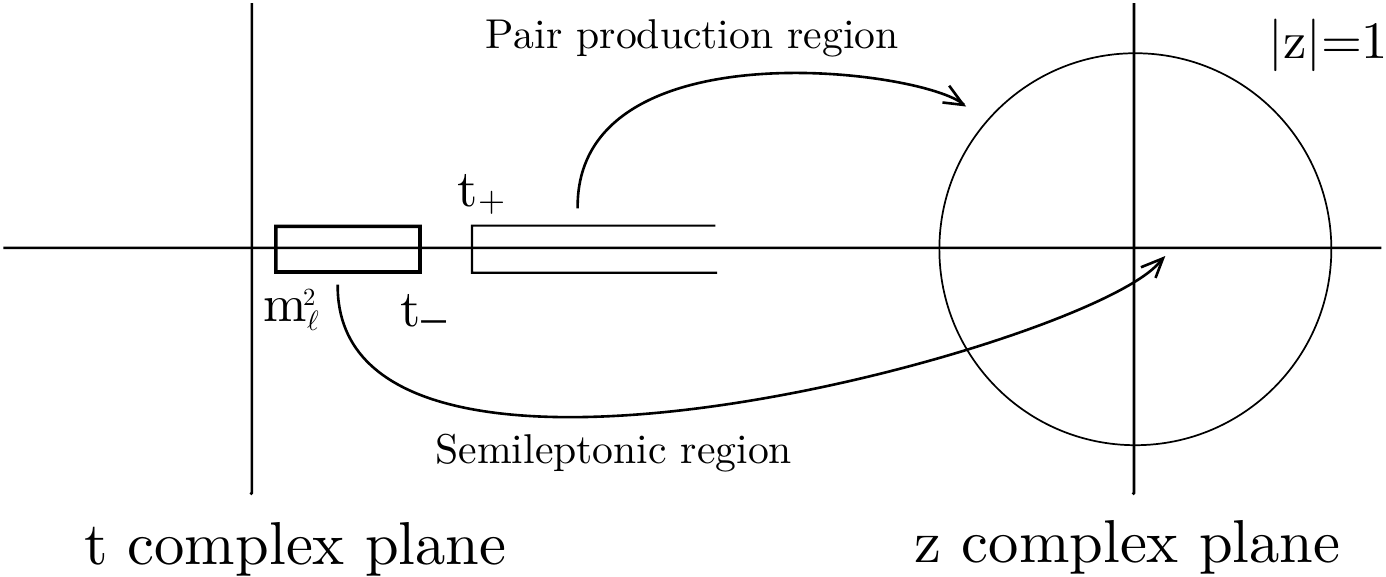}
  \caption{Graphic representation of the map between $t$ and $z$.}
  \label{tMapsToz}
  \end{center}                                                                                                                                                                  
\end{figure}

The analytical continuation is easily carried out after performing a conformal transformation $q^2\to z$ that maps the pair-production region onto the unit circle,
and the semileptonic region onto the real axis, as shown in Fig.~\ref{tMapsToz}.
A generic form factor $f$ can be written as a polynomial in $z$,
\begin{equation}
f = \frac{1}{\phi(z)B(z)}\sum_j a_j z^j
\end{equation}
where the outer functions $\phi(z)$ are calculated from the dispersion relation, the inner functions $B(z)$ or Blaschke factors take into account the contributing poles with
poorly known residues, and the coefficients $a_j$ are bounded by the \emph{weak unitarity constraints}, $\sum a_j^2 < 1$.
These are the fundamentals of the Boyd-Grinstein-Lebed (BGL) parametrization~\cite{Boyd:1995sq,Boyd:1995cf,Boyd:1997kz}, which is based only on very broad assumptions, like
analiticity of the form factors or unitarity of the theory, and thus it is model-independent.

However the most popular parametrization until very recently was the Caprini-Lellouch-Neubert (CLN) parametrization~\cite{Caprini:1997mu}.
Although it is based on the same theoretical grounds as the BGL parametrization, CLN has built in information about the relevant poles, as well as contributions from a variety
of channels that can help constrain the form factors even further, including cross symmetry between the four possible $B^{(\ast)}\to D^{(\ast)}\ell\nu$ semileptonic decays.
CLN also gives a closed expression for a main form factor $V_1$, and the rest $F_j$ are expressed as ratios $F_j/V_1$ using HQET up to $1/m_c$ order.

The CLN parametrization has a few properties that makes it quite appealing to the community: it offers simple, closed expressions for the form factors, instead of the complicated
outer functions of the BGL parametrization, it already has all the poles and susceptibilities built in, and its aggressive approximations give smaller errors for the form factor
fits than the general expressions offered by BGL.
Nonetheless, the CLN parametrization either does not include errors in the coefficients of the polynomial ansatz used to fit the form factors, or if it does, they are difficult
to implement, and most analysis ignored them. Moreover, the inputs employed in CLN are quite dated, and the approximations might be too aggressive.
The current consensus~\cite{Gambino:2020jvv} is that the CLN parametrization does not take into account certain systematic errors that are relevant at the level of precision we
have reached now, and thus we should abandon it in favor of the more general BGL parametrization.

An alternative to parametrizations are the dispersive matrix methods, introduced in~\cite{Lellouch:1995yv} and recently revived in~\cite{DiCarlo:2021dzg}.
But they are not broadly implemented by the community yet, and there are some pending theoretical issues in these methods that must be resolved\footnote{The
method has been widely discussed in the especialized workshop \emph{Challenges in semileptonic $B$ decays} in Barolo, 19$^{th}$-23$^{rd}$ April 2022. Unfortunately,
there are no written records of the discussions.}.

\subsection{Heavy quarks on the lattice}
Calculation of heavy-to-heavy decays on the lattice requires proper treatment for the heavy quarks.
This is a challenge, because the lattice spacings currently available are not fine enough to allow for a safe simulation of the bottom quark:
discretization errors grow as $\alpha^k(am_Q)^n$, which can easily get out of control for $am_Q \gtrsim 1$.
Ideally we would seek $am_b \ll 1$, but most state-of-the-art simulations are performed in the region $a\approx 0.12-0.045$ fm, which translates into $1.6-4.4$ GeV$^{-1}$,
leading to large discretization errors.

Earlier calculations mainly relied on effective actions for the heavy quarks.
The clear advantage of this approach is the possibility of directly simulating at the physical mass of the bottom quark.
On the downside, the effective theory requires a matching procedure to QCD which complicates the renormalization of the currents, introducing new systematic errors
that are not easy to take into account.

Latest simulations that use many ensembles at small enough values of the lattice spacing can afford a different approach: calculate the form factors at different (unphysical)
values of $m_b$, and extrapolate to the physical value at the end. The bottom quark mass has a large impact on the shape of the form factors, for it directly affects the available
energy for the leptons, and hence the kinematic range. Therefore, this approach is feasible when the extrapolation in $m_b$ is not large.
This method greatly simplifies the renormalization procedure, which can be done nonperturbatively, reducing an important source of systematic errors.

It is important to use the same regularization for the $b$ and the $c$ quark, because many systematic errors cancel out when computing the ratios
required to extract the form factors~\cite{Harrison:2017fmw}.

\section{Recent calculations}
Given the importance of the $B\to D^\ast\ell\nu$ channel for experiments (see Sec.~\ref{SecCT}), and the fact that no unquenched lattice calculations existed at nonzero recoil,
most efforts of the lattice community have been focused on filling this gap.
For that reason, the results for $B\to D^\ast\ell\nu$ dominate the most recent lattice calculations, with only a few mentions to $B\to D\ell\nu$.

\subsection{Fermilab - MILC $B\to D^\ast\ell\nu$}
\subsubsection{Lattice setup}
\begin{figure}[h]
  \begin{center}
  \includegraphics[width=0.5\textwidth,angle=0]{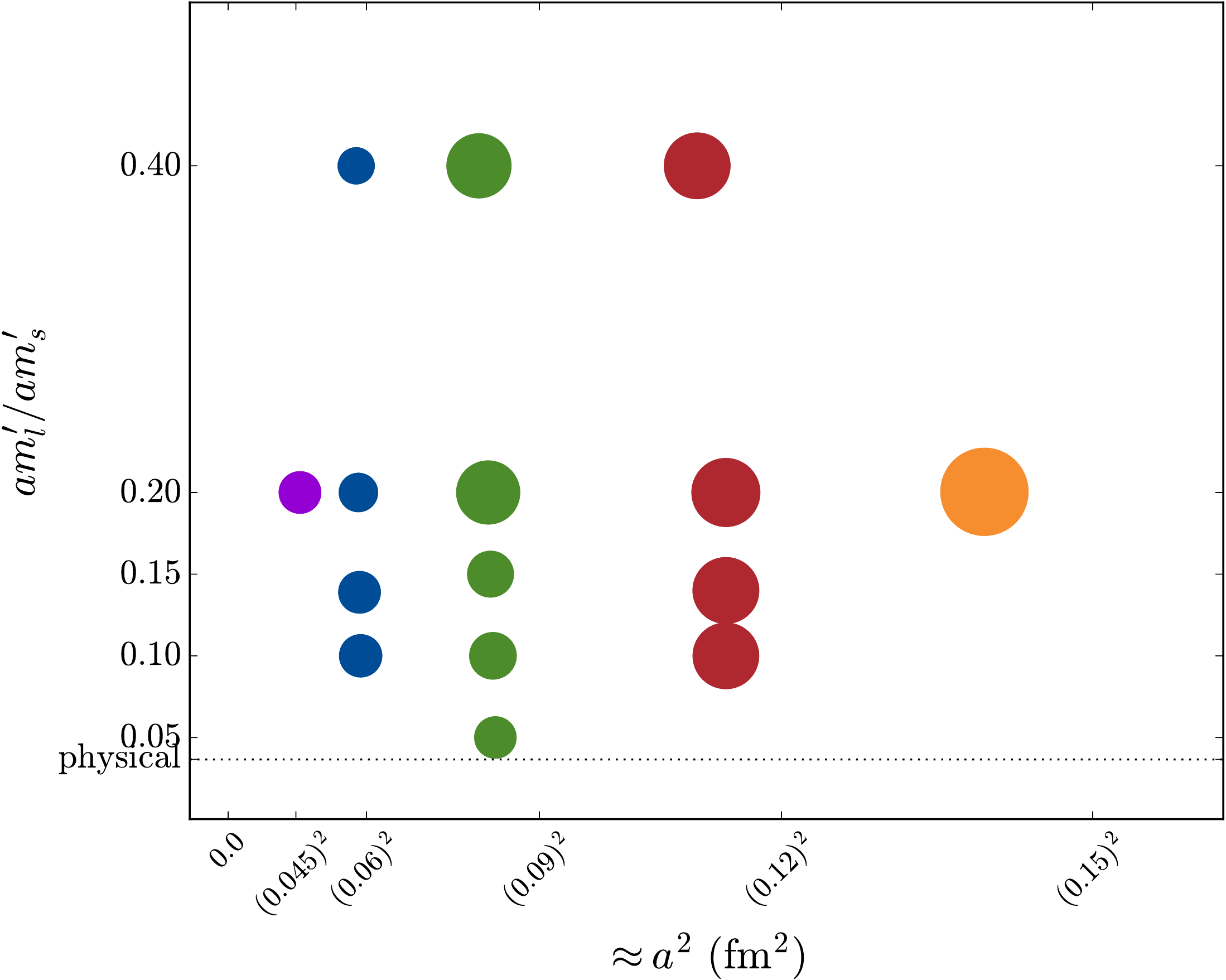}
  \caption{Ensembles employed in the Fermilab-MILC analysis. The area of each circle is proportional to the available statistics for the corresponding ensemble.}
  \label{FMEnsembles}
  \end{center}
\end{figure}

The most relevant calculation during the last year is the recently published Fermilab - MILC one~\cite{FermilabLattice:2021cdg}, that gives the first complete results for the form
factors of the $B\to D^\ast\ell\nu$ decay in the whole recoil range.
The calculation uses 15 ensembles of $N_f=2+1$ asqtad sea quarks, with the strange quark tuned to its physical mass.
Both heavy quarks, the $b$ and the $c$, use the Fermilab interpretation of the clover action, and their masses are tuned to their physical values through the $B_s$ and the $D_s$
mesons.
The ensembles differ in the values of their light quark mass and the lattice spacing, as shown in Fig.~\ref{FMEnsembles}.
The lightest pion mass reaches $m_\pi\approx 180$ MeV, which although light, makes the $D^\ast$ meson stable.

This analysis supersedes the previous one at zero recoil from Ref.~\cite{Bailey:2014tva} and takes a more conservative approach to deal with excited-states contamination.

\subsubsection{Published results}
The collaboration calculated the form factors at two different non-zero momenta for each ensemble, and the $h_{A_1}$ form factor also included data at zero recoil.
The chiral-continuun extrapolation is based on rooted staggered chiral perturbation theory (rS$\chi$PT), with some additions from heavy quark effective theory (HQET) to
deal with heavy-light observables. A combined fit of all the data, including correlations, is performed, with a resulting $\chi^2/\textrm{dof}=85.2/95$. 
There is an increase in the error of $h_{A_1} = 0.909(17)$ from the result in Ref.~\cite{Bailey:2014tva}, $h_{A_1} = 0.906(13)$, due to the more conservative treatment of the
excited states, but the agreement with previous results, including HPQCD's Ref.~\cite{Harrison:2017fmw} is excellent.

\begin{figure}[h]
  \begin{center}                                                                                                                                                                  
  \includegraphics[width=0.45\textwidth,angle=0]{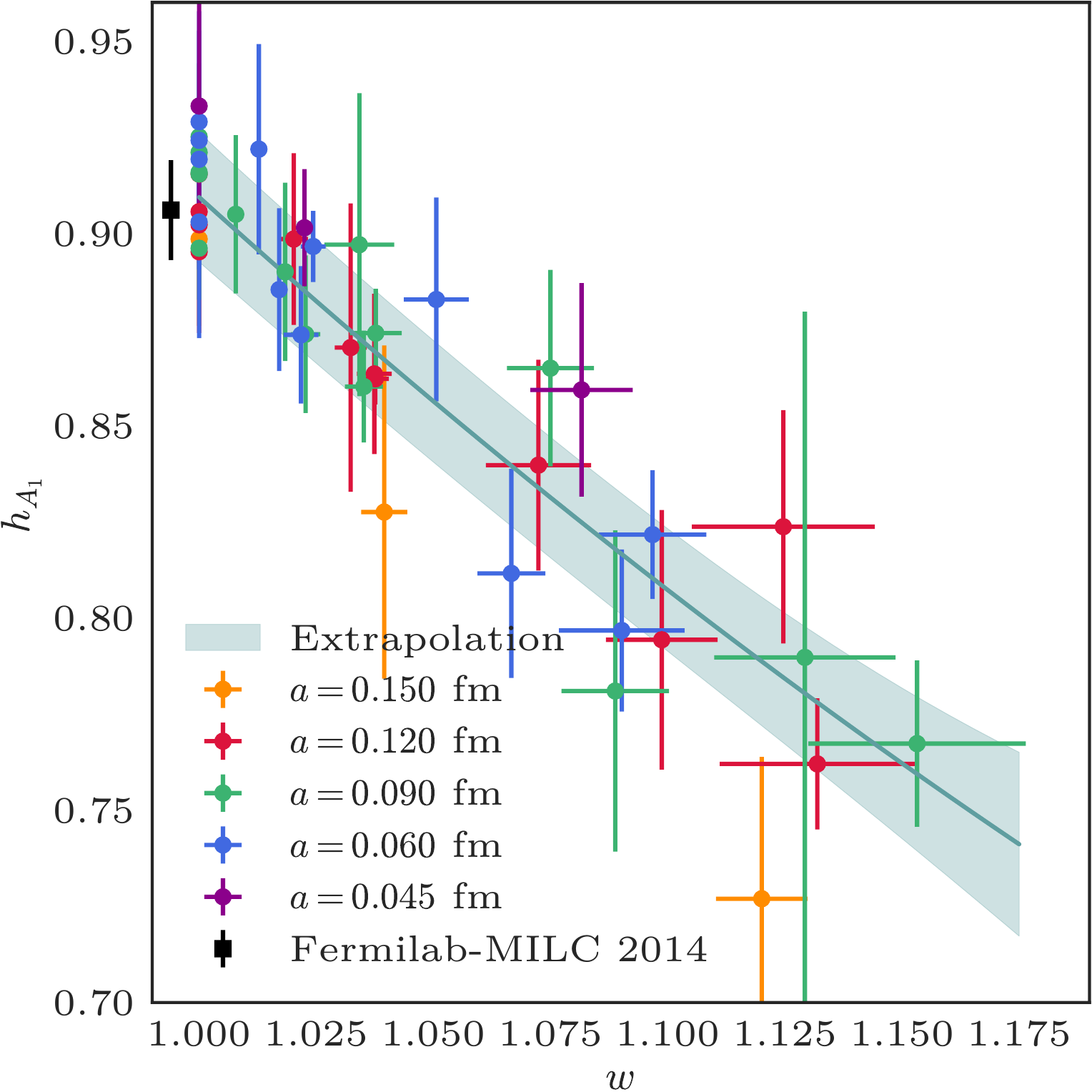} \hfill
  \includegraphics[width=0.45\textwidth,angle=0]{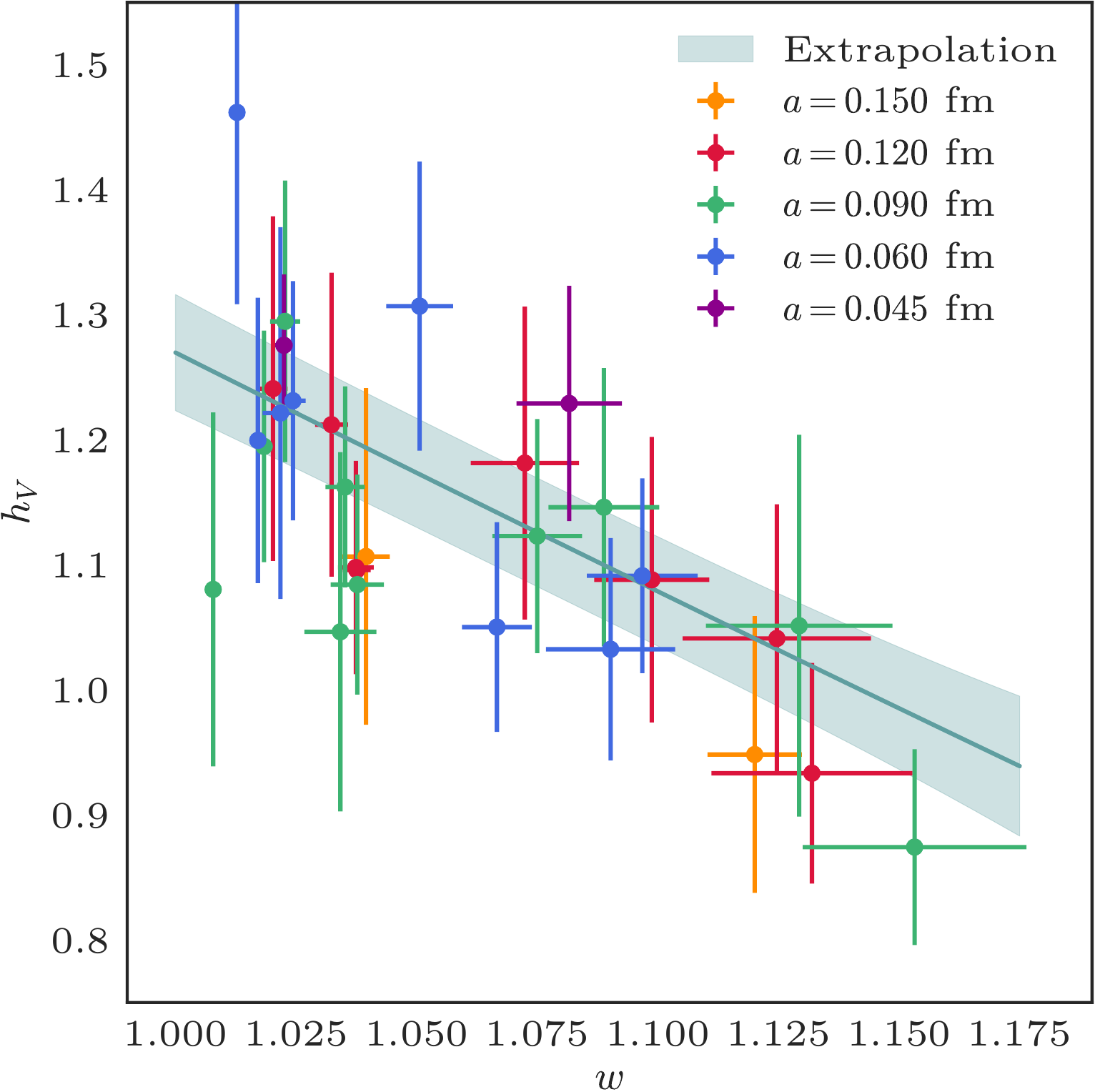}
  \caption{Chiral-continuum extrapolation of the Fermilab-MILC data for the $h_{A_1}$ (left) and the $h_V$ (right) form factors.}
  \label{FMA1V}
  \end{center}                                                                                                                                                                  
\end{figure}

\begin{figure}[h]
  \begin{center}                                                                                                                                                                  
  \includegraphics[width=0.45\textwidth,angle=0]{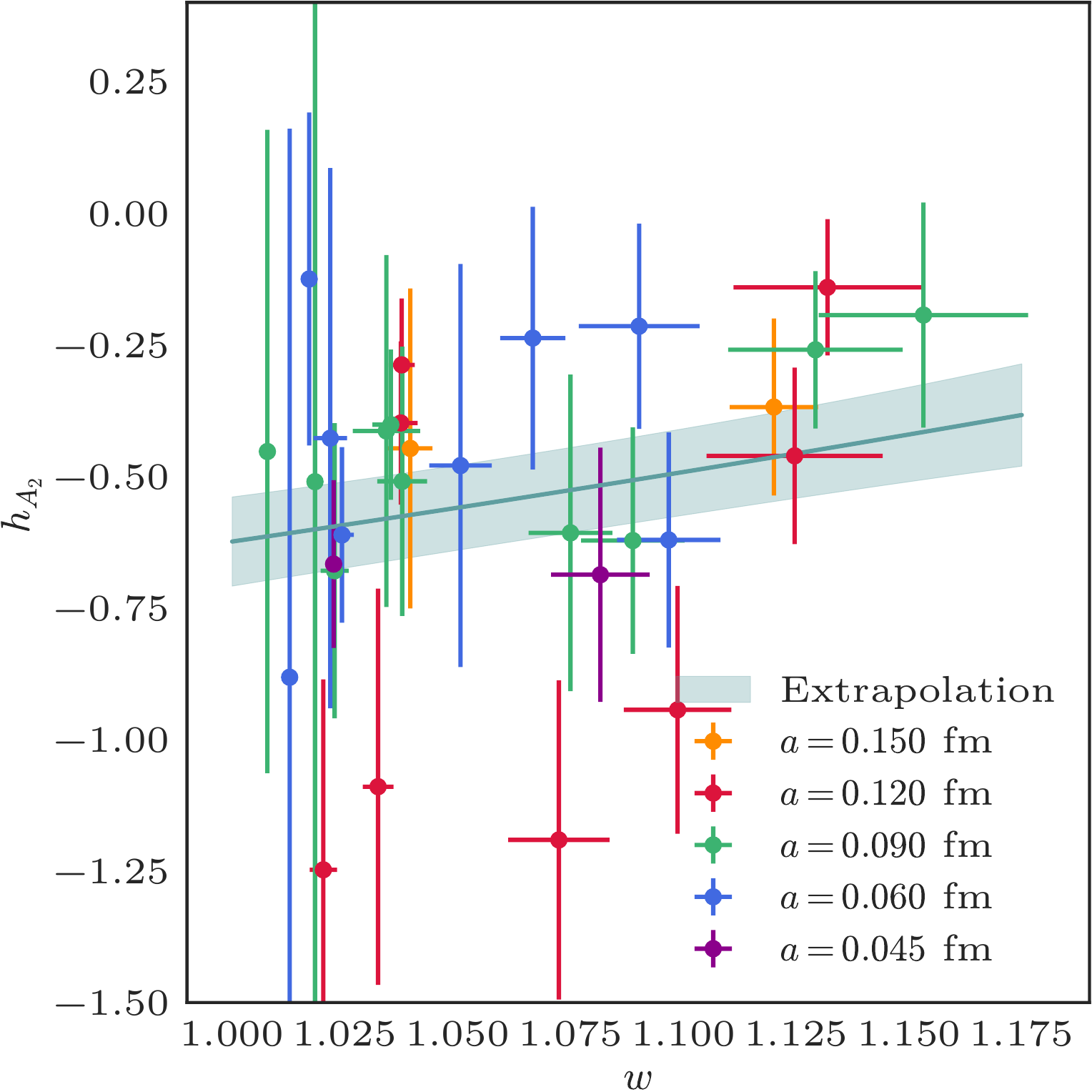} \hfill
  \includegraphics[width=0.45\textwidth,angle=0]{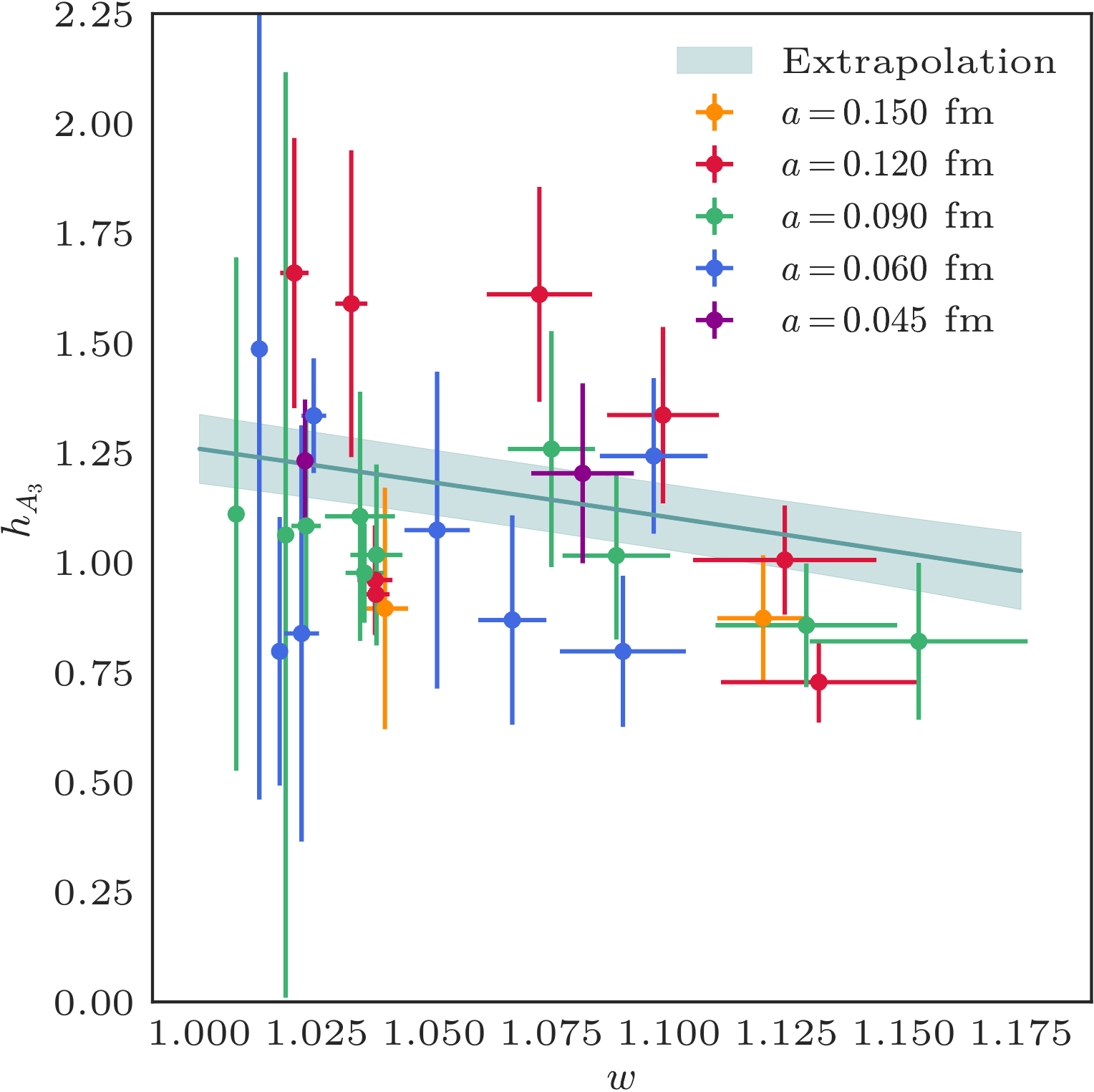}
  \caption{Chiral-continuum extrapolation of the Fermilab-MILC data for the $h_{A_2}$ (left) and the $h_{A_3}$ (right) form factors.}
  \label{FMA23}
  \end{center}                                                                                                                                                                  
\end{figure}

\begin{figure}[h]
  \begin{center}                                                                                                                                                                  
  \includegraphics[width=0.45\textwidth,angle=0]{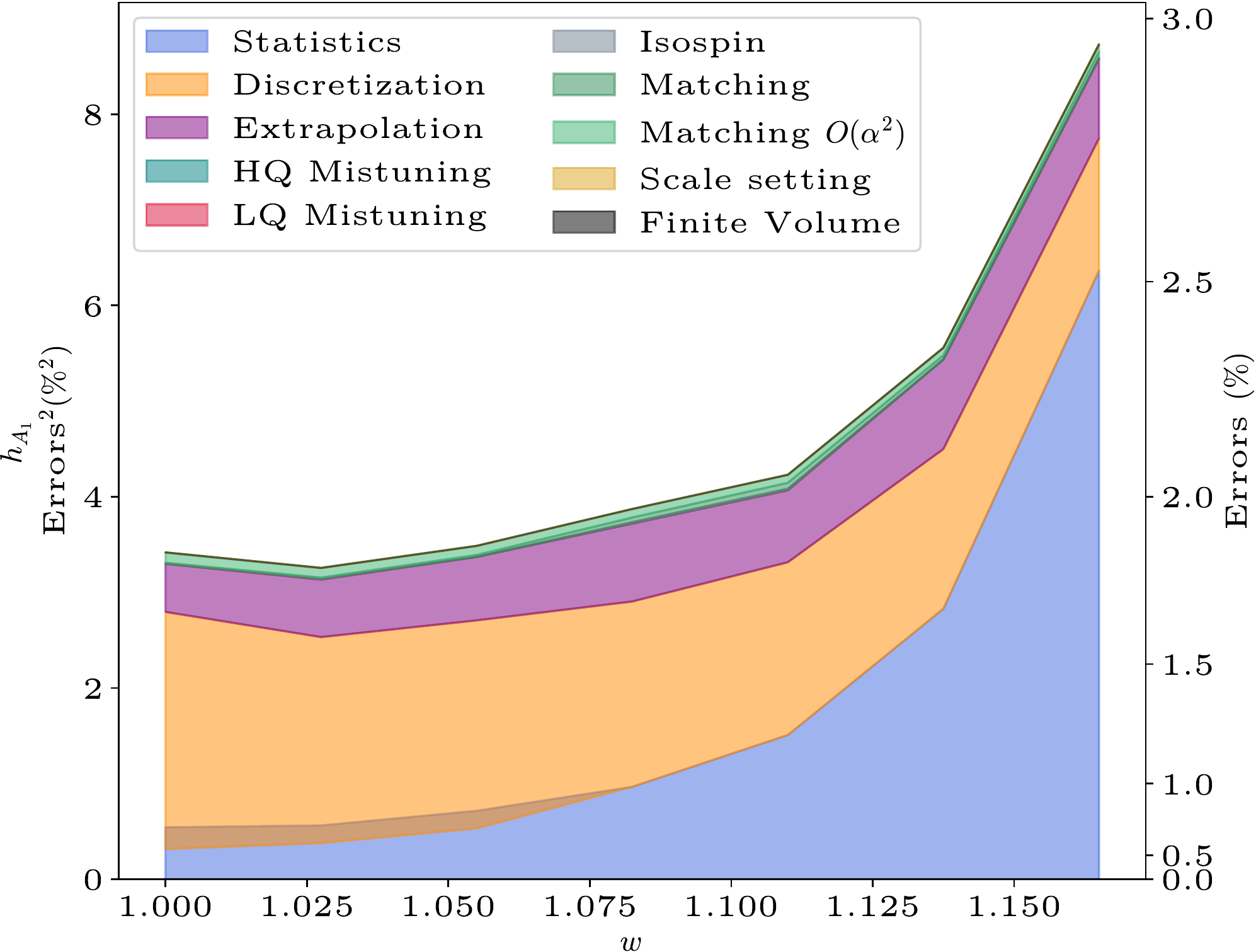} \hfill
  \includegraphics[width=0.45\textwidth,angle=0]{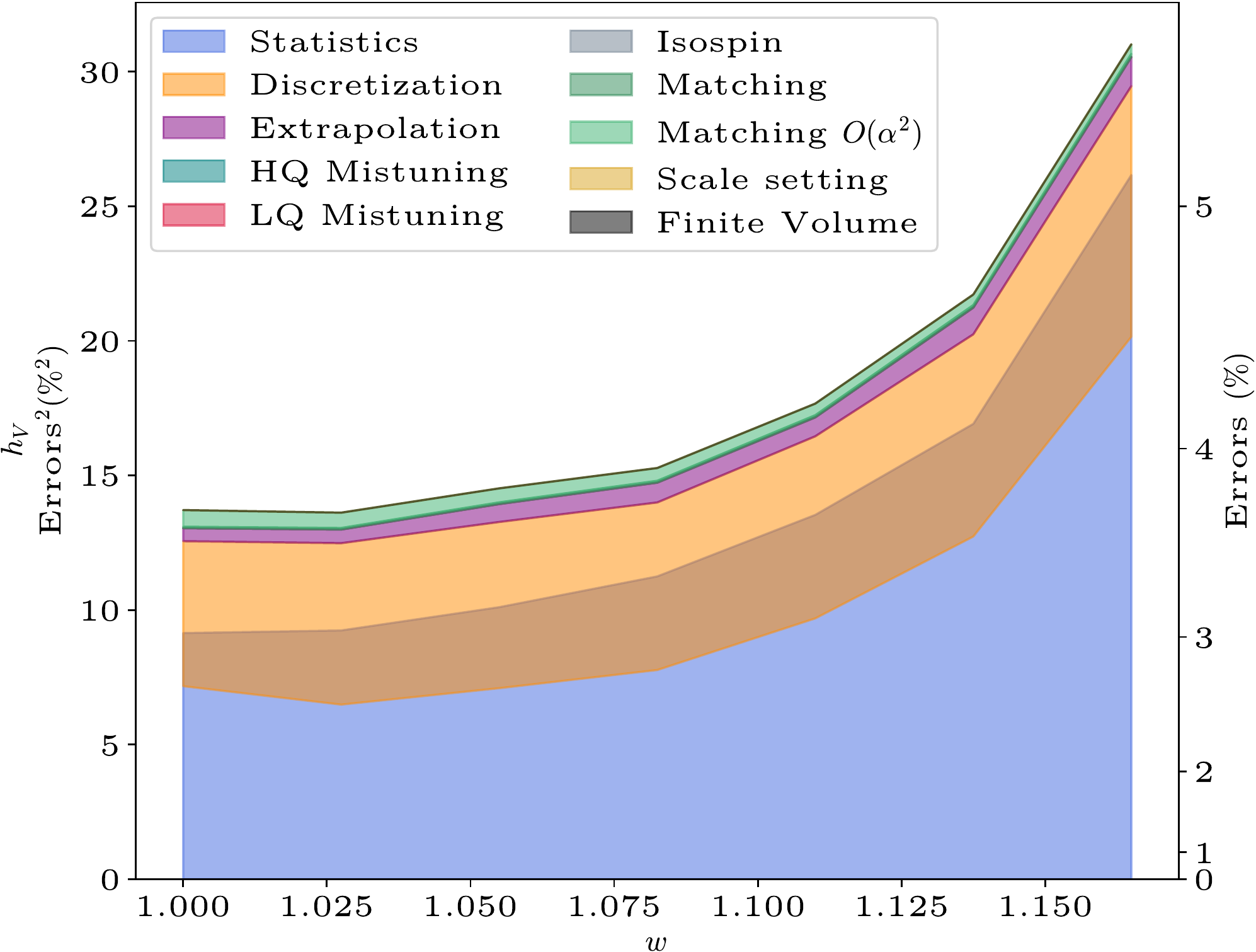}
  \caption{Complete error budget for the Fermilab-MILC $h_{A_1}$ (left) and $h_V$ (right) form factors.}
  \label{FMEBA1V}
  \end{center}                                                                                                                                                                  
\end{figure}

\begin{figure}[h]
  \begin{center}                                                                                                                                                                  
  \includegraphics[width=0.45\textwidth,angle=0]{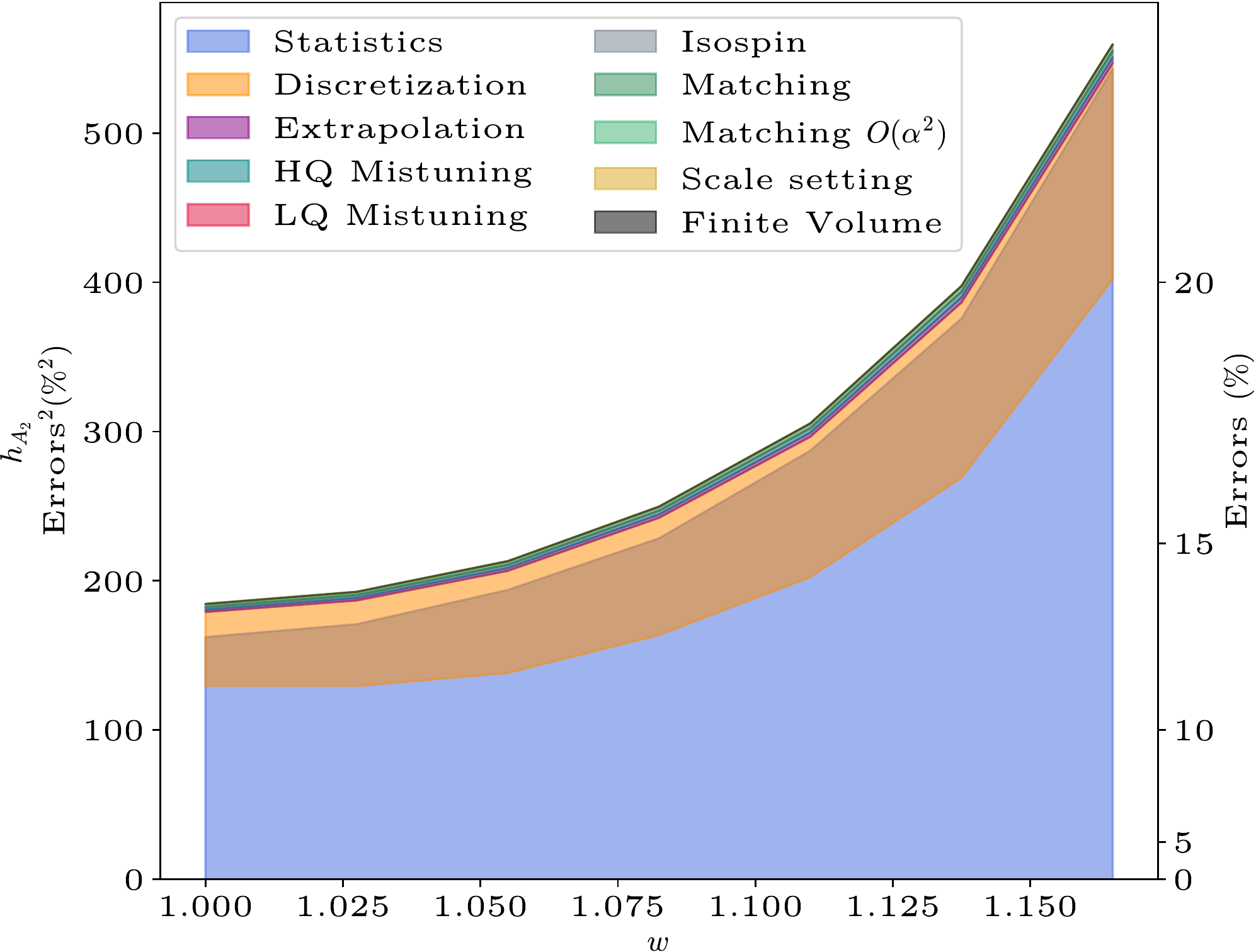} \hfill
  \includegraphics[width=0.45\textwidth,angle=0]{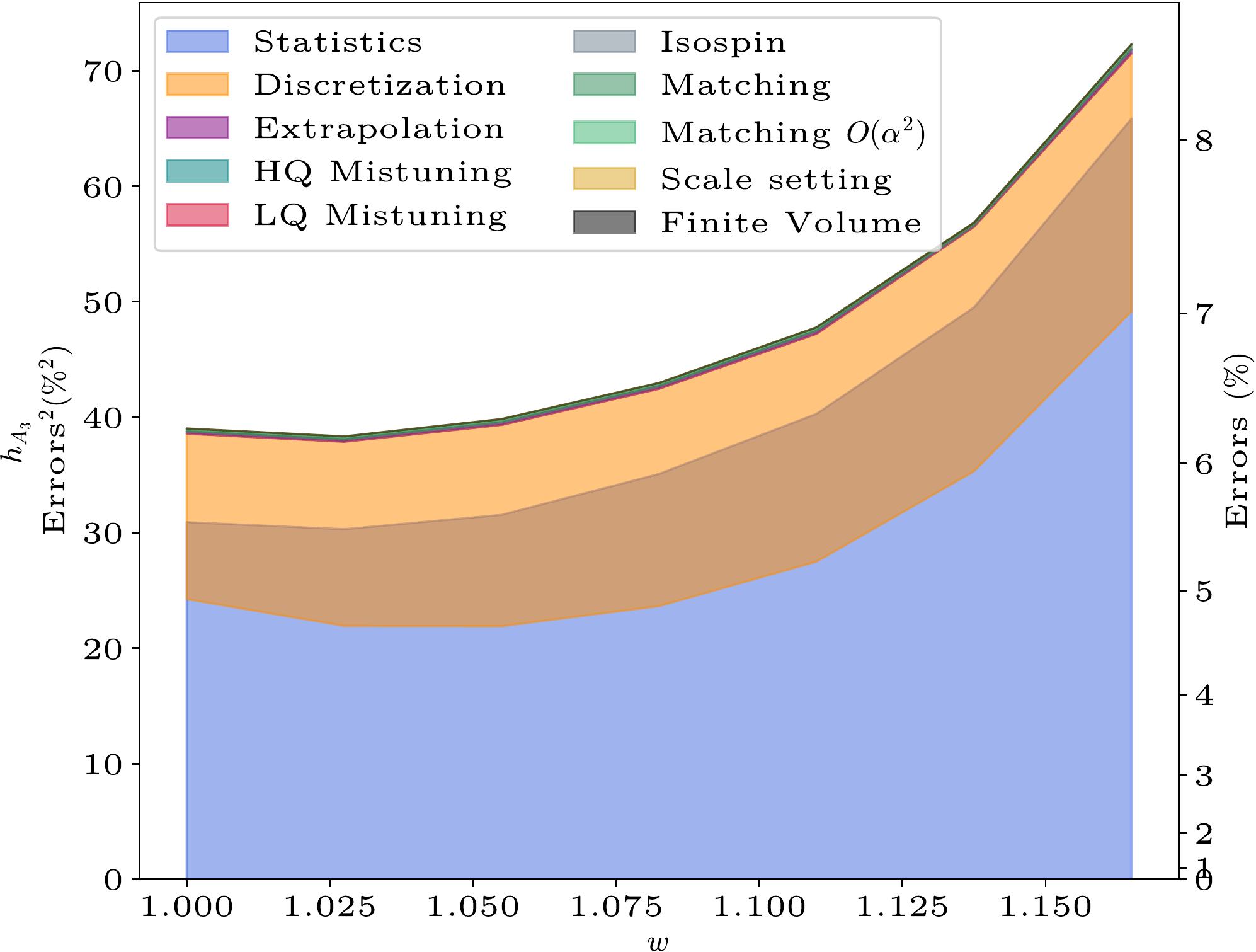}
  \caption{Complete error budget for the Fermilab-MILC $h_{A_2}$ (left) and $h_{A_3}$ (right) form factors.}
  \label{FMEBA23}
  \end{center}                                                                                                                                                                  
\end{figure}

Results for all the form factors are shown in Figs.~\ref{FMA1V} and~\ref{FMA23}. A careful analysis of systematic errors is available, and can be
checked in Figs.~\ref{FMEBA1V} and~\ref{FMEBA23}. The main contributions come from statistics and discretization errors, and the collaboration is working on ways to improve
the precision in both areas in future analyses.

From the chiral-continuum extrapolation results, synthetic data is generated for the form factors, and the BGL parametrization is used to extend the validity of the data
to the whole recoil range. The shape of the decay amplitude $\mathcal{F}(w)$ can be compared with experiment if the data is properly normalized. This comparison is shown in
the left pane of Fig.~\ref{DecayComp}. The LQCD curve consistently stays below the experimental curves, but the differences are within $2\sigma$. More worrisome is the difference
in the slopes of the decay amplitude at small recoil: since the lattice synthetic data is highly correlated, the slope has a much smaller error than what one would expect from
the plot. It is clear that there is an incompatibility with the shapes of the decay amplitude, and further analysis is needed to shed some light on the issue.

Assuming the differences in shapes are due to statistical fluctuations, one can extract $|V_{cb}|$ using Belle~\cite{Waheed:2018djm} and BaBar~\cite{Dey:2019bgc} data.
Belle provides unfolded data, efficiency matrices and correlations, as well as a procedure to properly analyze the data in their publication.
There is some confusion regarding the statistical correlation matrix, which is missing essential features that should appear by construcion (see Ref.~\cite{Bobeth:2021lya}),
but the collaboration uses Belle data as it is. BaBar provides a BGL fit from which synthetic data can be extracted.

\begin{table}
  \caption{Quality of fit for the different fits attempted by the Fermilab-MILC collaboration. The last column uses lattice QCD data only for renormalization.}
  \label{CompResults}                                                                                                                
  \begin{tabular}{rccccc}
    \hline                                                                                                                          
                    & {Lattice QCD} & {Lattice + BaBar} & {Lattice + Belle} & {Lattice + both} & {$h_{A_1}(1)$ + both} \\
    \hline                                                                                                                          
$\chi^2/\text{dof}$ &    {0.63/1}   &     {8.50/4}      &     {111/79}      &     {126/84}     &        {104/76}       \\
    \hline
  \end{tabular}
\end{table}

Table~\ref{CompResults} shows the deaugmented $\chi^2/\textrm{dof}$ of a variety of BGL fits the collaboration attempted, but a few points must
be highlighted: First, the BGL fit to LQCD data using quadratic expansions for all the form factors gives and excellent value of $\chi^2/\textrm{dof}$; second, any fit that
includes Belle data displays a $\chi^2/\textrm{dof}$ higher than one. Even a combined fit using only experimental input results in a relatively low quality of fit, suggesting
there are some tensions between experimental datasets. When including LQCD in the mix, the quality of fit decreases even further to $\chi^2/\textrm{dof}\approx 1.5$, clearly showing
the existing differences in the shapes of the decay amplitude. In this context, the extracted value of $|V_{cb}| = 38.40(78)\times 10^{-3}$, which includes the Coulomb correction,
applied to both Belle and BaBar datasets, is perfectly compatible with previous determinations, and does not answer the question of why the inclusive and the exclusive determinations
differ so much.

An $R(D^\ast)$ calculation is straightforward, and the collaboration provides two values: one extracted solely from LQCD data, and hence a pure theoretical determination, giving
$R(D^\ast)_{\textrm{Lat}} = 0.265(13)$; and another one calculated from the combined fit with experiment, $R(D^\ast)_{\textrm{Lat+Exp}} = 0.2484(13)$. The latter assumes new physics
are only visible at higher lepton masses, i.e., for the $\tau$, and the former is compatible with the HFLAV result $R(D^\ast) = 0.295(14)$ within $1.6\sigma$.
Fig.~\ref{RDstComp} shows a comparison of the different $R(D^\ast)$ results.

A newer analysis of $B\to D^{(\ast)}\ell\nu$ that features the HISQ action for the light sector and the Fermilab action for the heavy sector is currently in the works. 
This newer analysis promised an $R(D)-R(D^\ast)$ correlated result, and is beng done in coordination with a $B\to\pi\ell\nu$ analysis, in order to provide a $|V_{ub}|-|V_{cb}|$
correlated result as well.
Also, results for an all-HISQ analysis of the $B\to D\ell\nu$ form factors by the Fermilab-MILC collaboration have been reported in this conference, see Ref.~\cite{Lytle:2022ps}.

\subsection{JLQCD $B\to D^{(\ast)}\ell\nu$}
\subsubsection{Lattice setup}
The JLQCD collaboration is on the verge of finishing their own computation of the $B\to D^\ast\ell\nu$ form factors using a very different setup from that of their Fermilab-MILC colleagues.
This provides a good crosscheck of results, since the systematic errors are very different.

In this calculation, 8 ensembles of $N_f=2+1$ Domain Wall sea quarks are employed. The strange quark is always tuned to its physical mass, but the light quarks are heavier than physical,
resulting in values of $m_\pi$ ranging from $\approx 500$ MeV down to $\approx 230$ MeV, ensuring that the $D^\ast$ meson is stable.
The heavy quarks are treated relativistically, using the same Domain Wall action than their light counterparts. Hence, the renormalization of currents becomes non-perturbative, and
it is straightforward and more precise, but the method requires an extrapolation in $m_b$ to reach physical results. Fig.~\ref{JLEnFFs} shows the distribution of the different ensembles
employed in this calculation.

\subsubsection{Preliminary results}
\begin{figure}[h]
  \begin{center}                                                                                                                                                                  
  \includegraphics[width=0.23\textwidth,angle=0]{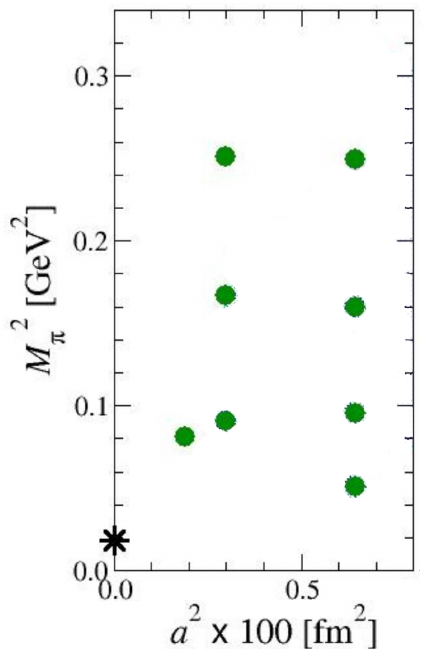} \hfill
  \includegraphics[width=0.73\textwidth,angle=0]{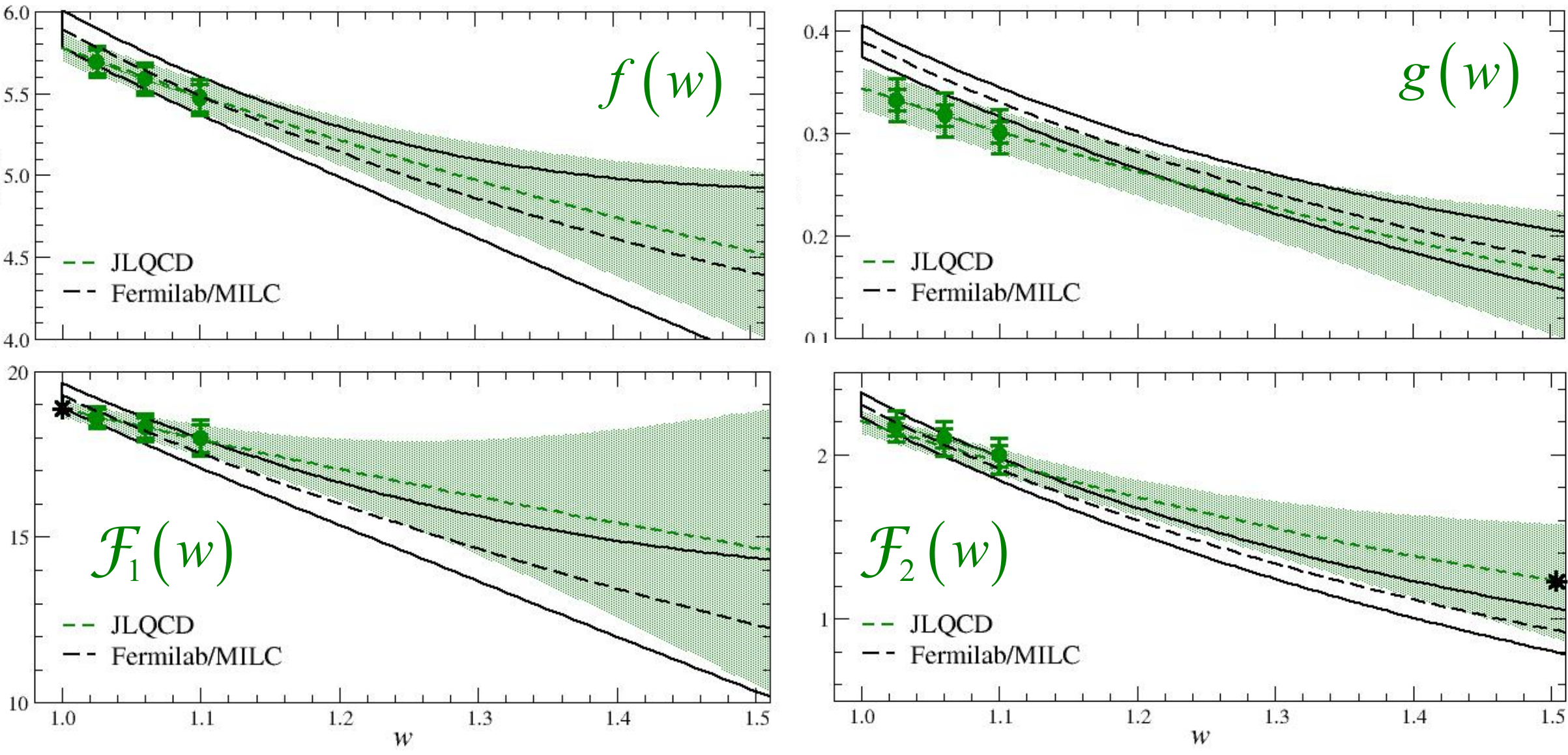}
  \caption{Left: Ensembles employed in the JLQCD analysis. Right: JLQCD preliminary form factor data, compared with Fermilab-MILC published results.}
  \label{JLEnFFs}
  \end{center}                                                                                                                                                                  
\end{figure}

\begin{figure}[h]
  \begin{center}                                                                                                                                                                  
  \includegraphics[width=0.57\textwidth,angle=0]{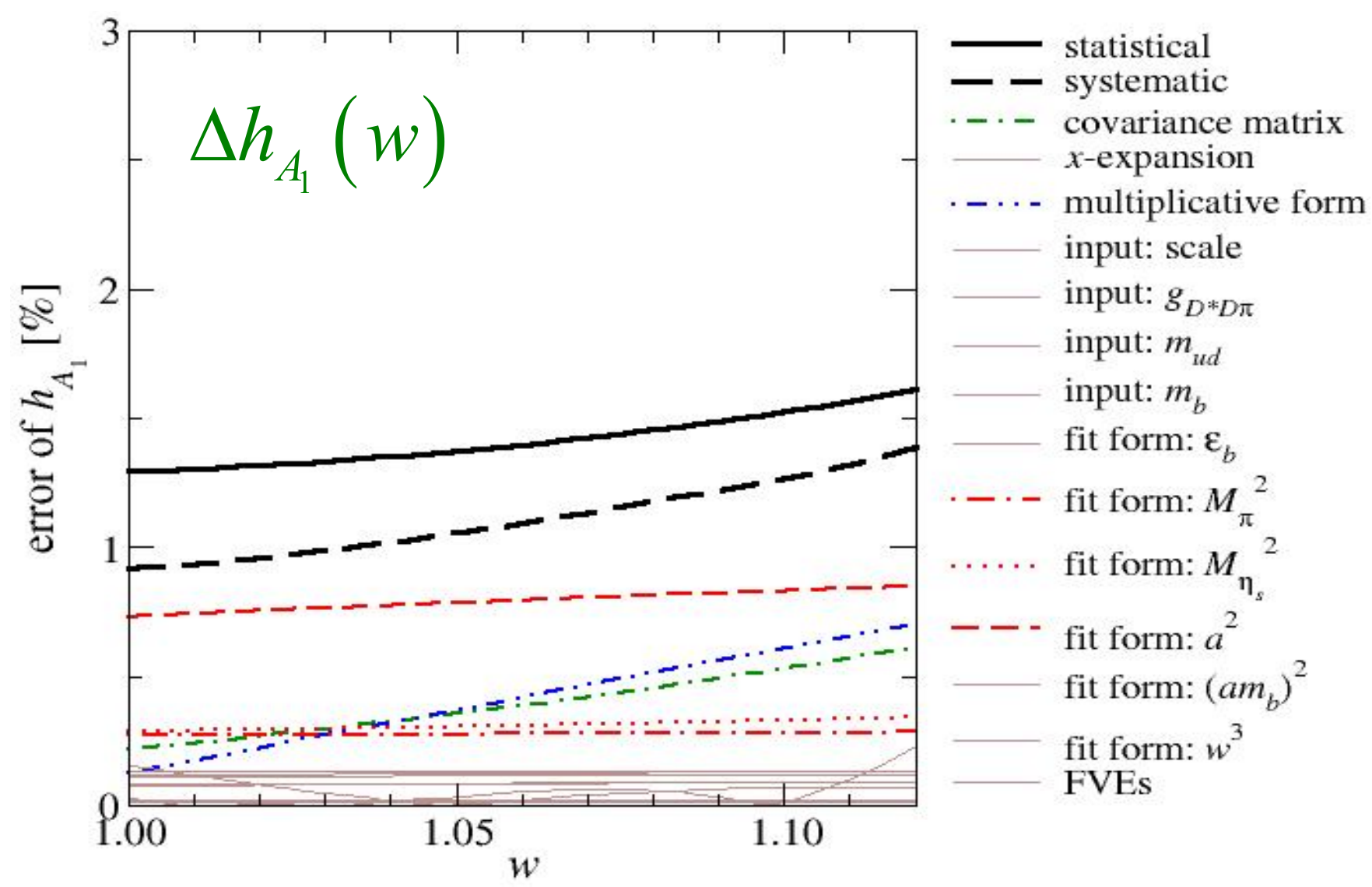} \hfill
  \includegraphics[width=0.38\textwidth,angle=0]{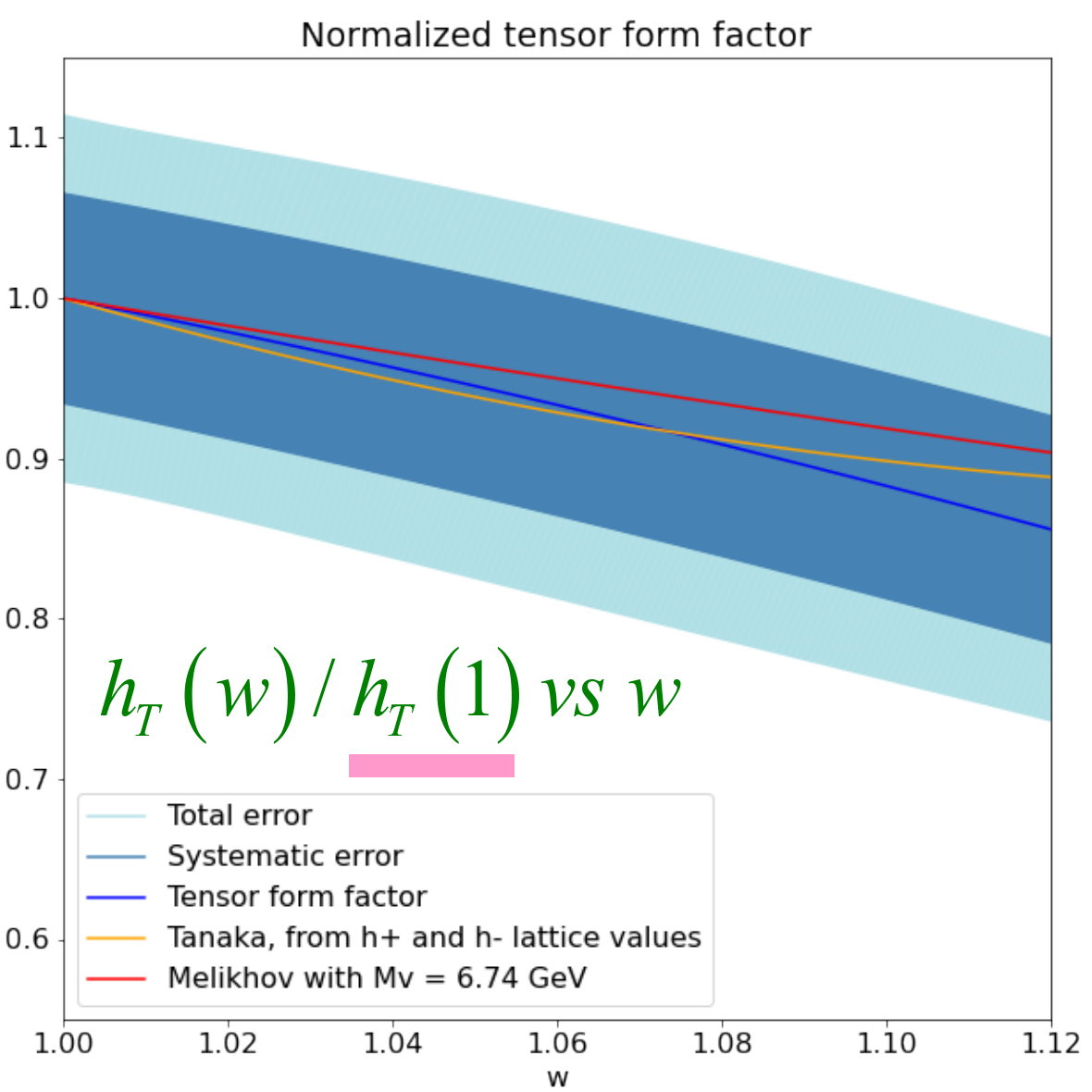}
  \caption{Left: preliminary error budget for the JLQCD form factor $h_{A_1}$. The other form factors show a similar error budget. Right: JLQCD preliminary results for the $h_T$ tensor form factor.}
  \label{JLSysT}
  \end{center}                                                                                                                                                                  
\end{figure}

Figure~\ref{JLEnFFs} shows the preliminary results of the form factors, compared with the published results of the Fermilab-MILC analysis. The figure compares the form factors in the BGL basis,
$f$, $g$, $\mathcal{F}_1$ and $\mathcal{F}_2$, which are functions of the $h_X$. In general, there is a good agreement between both
analyses, displaying a large overlap of form factors over the whole recoil region, but the slopes obtained by JLQCD are milder. There is a preliminary, but detailed error budget
for each form factor, and the left pane of Fig.~\ref{JLSysT} shows a representative result: as in the Fermilab-MILC case, the largest contribution to the error comes from statistics, followed by the
systematics associated with discretization errors.

Their synthetic data points are plotted in Fig.~\ref{DecayComp}, along with the results of Fermilab-MILC and the experimental curves. The JLQCD data points fit nicely in the gaps left by the experimental
points, and one can notice by eye that the compatibility of JLQCD data and experimental data is higher than in the Fermilab-MILC case, and a combined fit using Belle and JLQCD input gives
a $\chi^2/\textrm{dof} = 0.94$. Surprisingly, the differences in the results for $|V_{cb}|$ and $R(D^\ast)$ between both collaborations are not very large: JLQCD reports the preliminary result
$|V_{cb}| = 39.85(95)\times 10^{-3}$ for a fit including Belle and LQCD data, without including the Coulomb factor, and the equivalent fit using Fermilab-MILC data results in
$|V_{cb}| = 38.60(86)\times 10^{-3}$, For $R(D^\ast)$ the preliminary result shown by JLQCD is $0.253(22)$, with large enough errors to make it compatible with both the HFLAV average and
previous theoretical determinations (see Fig.~\ref{RDstComp}).

The JLQCD is also carrying out a $B\to D\ell\nu$ analysis, including form factors relevant for BSM calculations. Their preliminary result for the tensor form factor $h_T$ is missing the
normalization, but its shape is compatible with earlier theoretical predictions, as shown in the right pane of Fig.~\ref{JLSysT}. 

\subsection{HPQCD $B\to D^\ast\ell\nu$}
\subsubsection{Lattice setup}
\begin{figure}[h]
  \begin{center}
  \includegraphics[width=0.50\textwidth,angle=0]{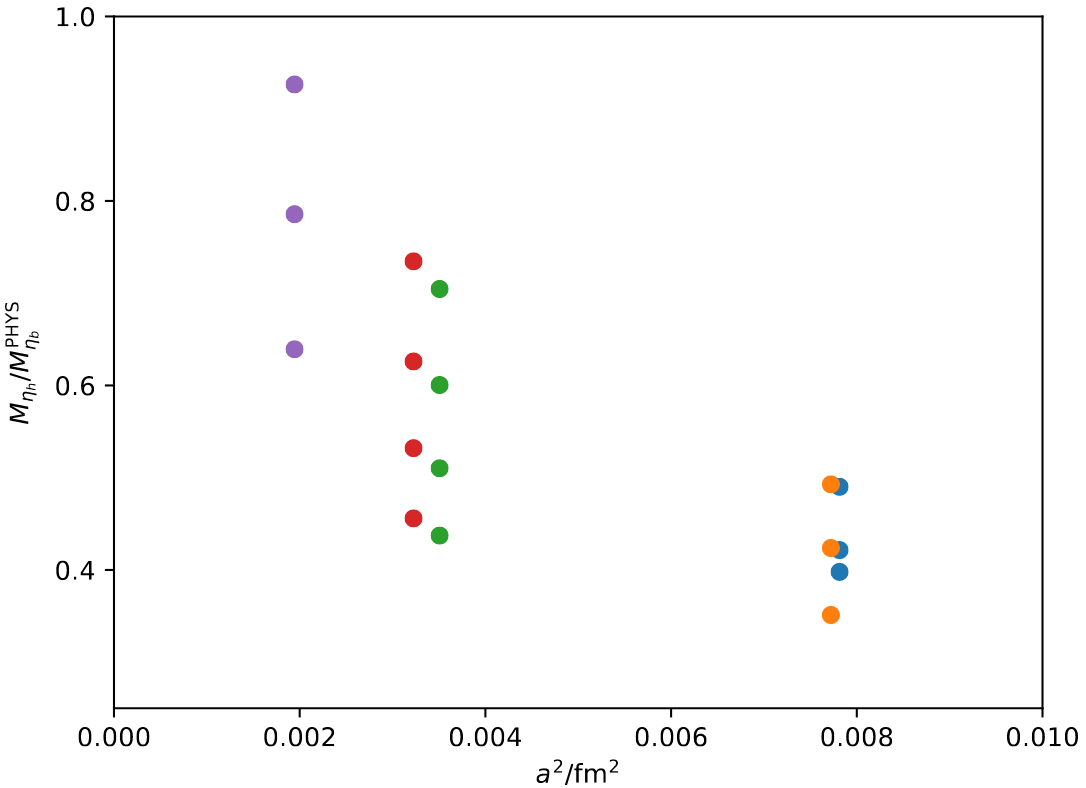}
  \caption{Ensembles employed in the HPQCD analysis. The different points with the same $x$ coordinate represent calculations with different values of $m_b$ on the same ensemble.}
  \label{HQEnsembles}
  \end{center}
\end{figure}

The analysis of the HPQCD collaboration employs five different ensembles of $N_f=2+1+1$ sea HISQ quarks. The $b$ quark also uses the HISQ action, but at unphysical masses, and an extrapolation
in $m_b$ is required to extract physical results. The pion masses range from $330$ MeV down to $129$ MeV, and the instability of the $D^\ast$ meson is dealt with by using chiral perturbation
theory. As a difference with the Fermilab-MILC analysis, correlators are available at a variety of momenta, which allows a reliable chiral-continuum extrapolation of the form factors covering
the whole recoil range. A summary of the ensembles and the data available for this analysis is gathered in Fig.~\ref{HQEnsembles}.

\subsubsection{Preliminary results}
\begin{figure}[h]
  \begin{center}                                                                                                                                                                  
  \includegraphics[width=0.45\textwidth,angle=0]{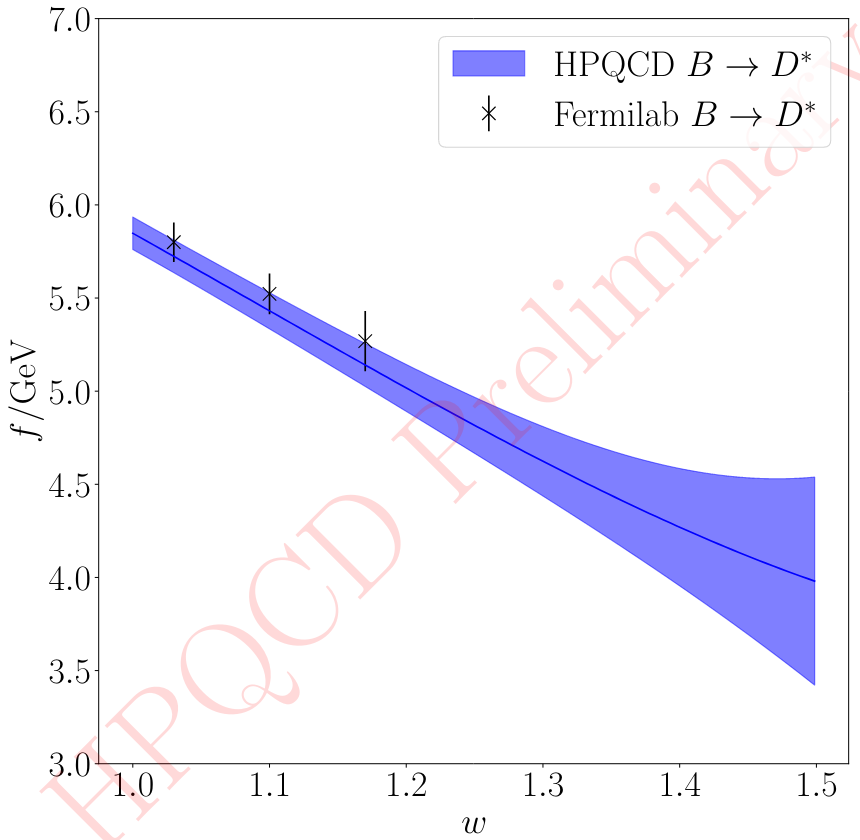} \hfill
  \includegraphics[width=0.45\textwidth,angle=0]{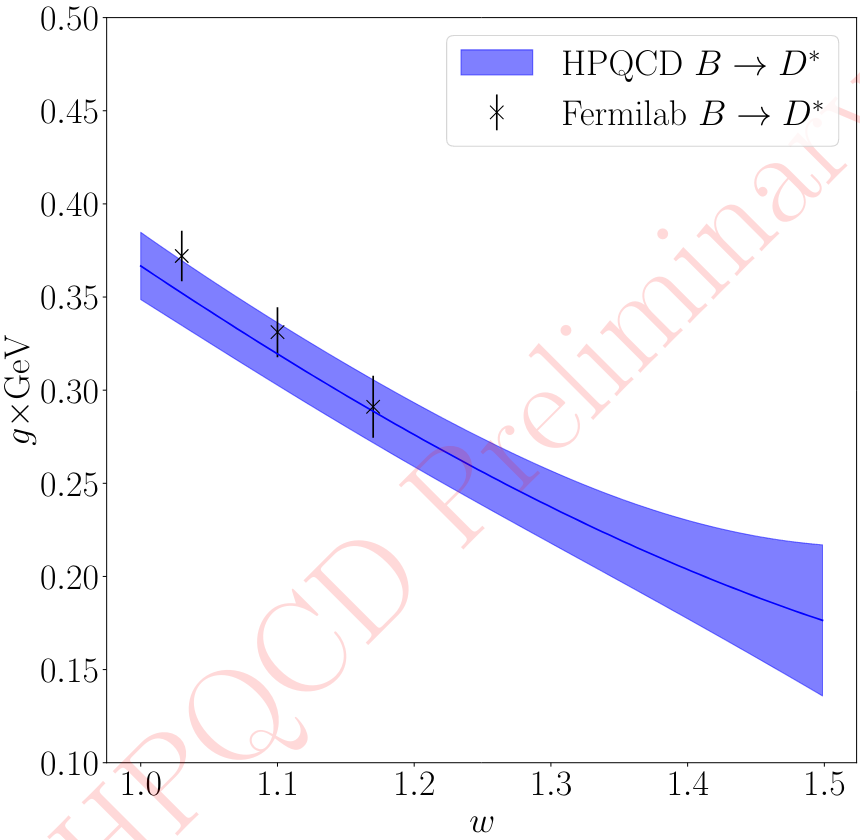}
  \caption{Chiral-continuum extrapolation of the HPQCD data for the $f$ (left) and the $g$ (right) form factors, compared with the Fermilab-MILC published results.}
  \label{HP-FF1}
  \end{center}                                                                                                                                                                  
\end{figure}

\begin{figure}[h]
  \begin{center}                                                                                                                                                                  
  \includegraphics[width=0.45\textwidth,angle=0]{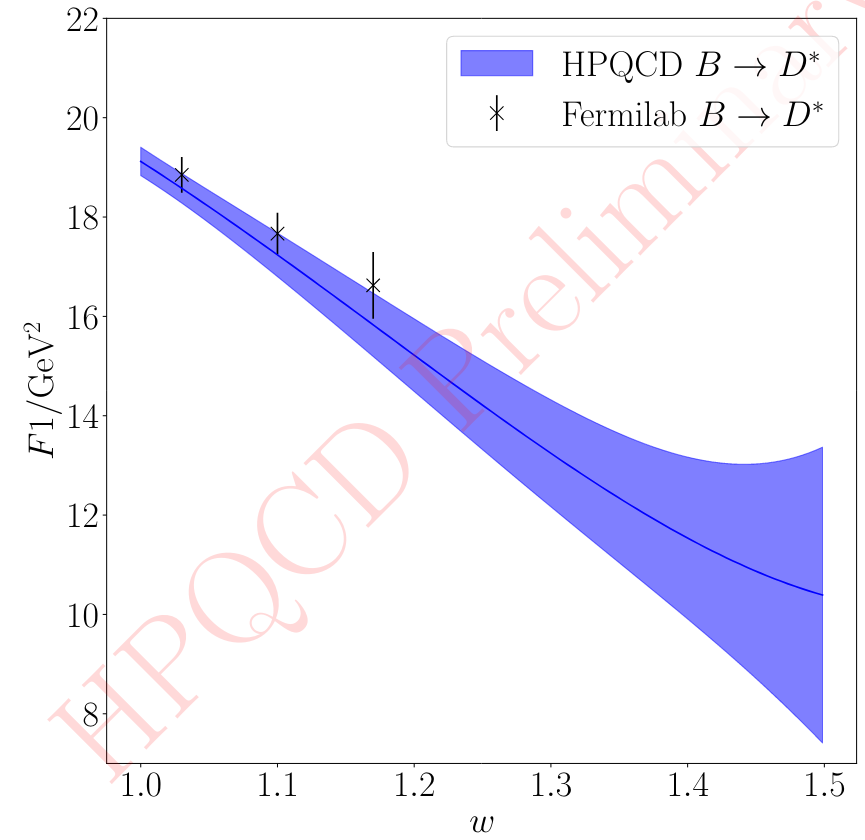} \hfill
  \includegraphics[width=0.45\textwidth,angle=0]{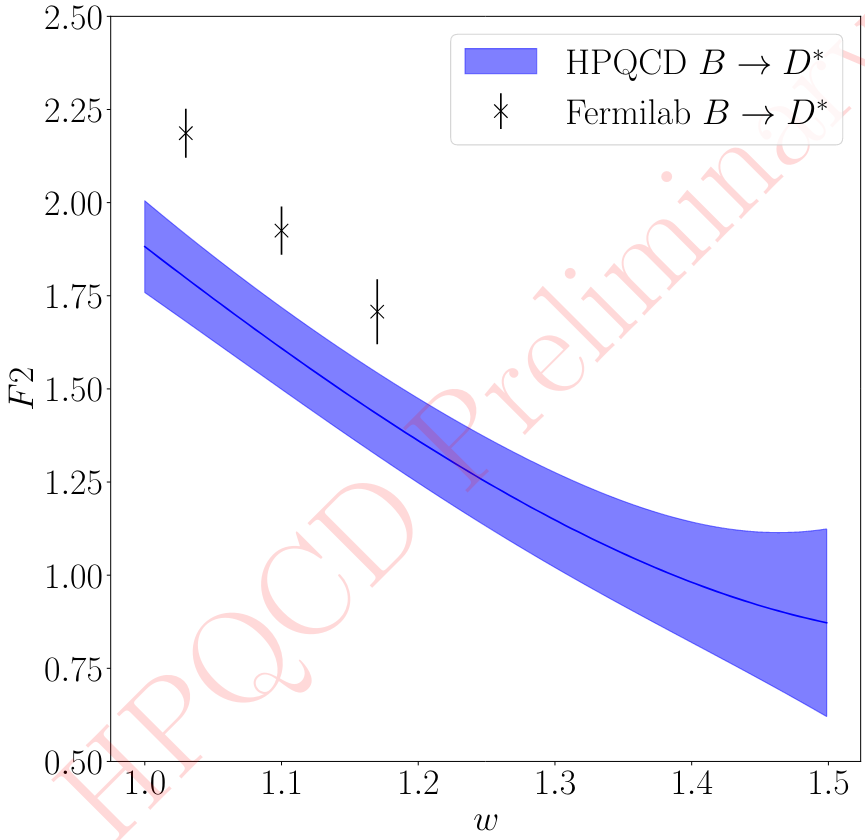}
  \caption{Chiral-continuum extrapolation of the HPQCD data for the $\mathcal{F}_1$ (left) and the $\mathcal{F}_2$ (right) form factors, compared with the Fermilab-MILC published results.}
  \label{HP-FF2}
  \end{center}                                                                                                                                                                  
\end{figure}

The lattice data is fitted to a chiral-continuum extrapolation that takes both the light and the $b$ quark masses to their physical values, giving results in the full recoil range for
the four form factors. Figures~\ref{HP-FF1} and~\ref{HP-FF2} compare the HPQCD results with the Fermilab-MILC synthetic data, showing a good agreement in $g$, $f$ and $\mathcal{F}_1$.
The differences are larger for $\mathcal{F}_2$, which shows an obvious disagreement.
%CAMBIAR A f,g,F1,F2

\begin{figure}[h]
  \begin{center}                                                                                                                                                                  
  \includegraphics[width=0.46\textwidth,angle=0]{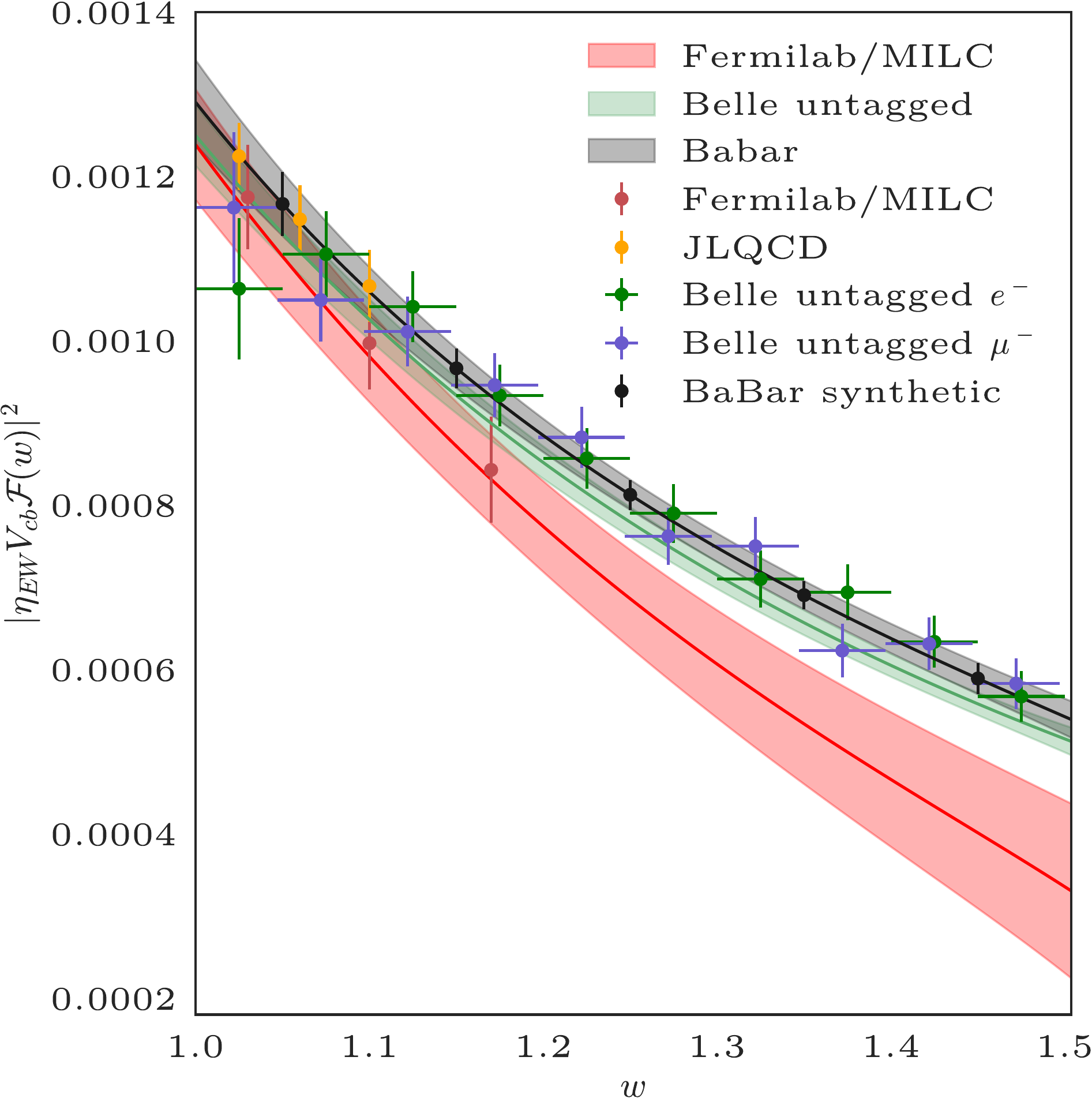} \hfill
  \includegraphics[width=0.49\textwidth,angle=0]{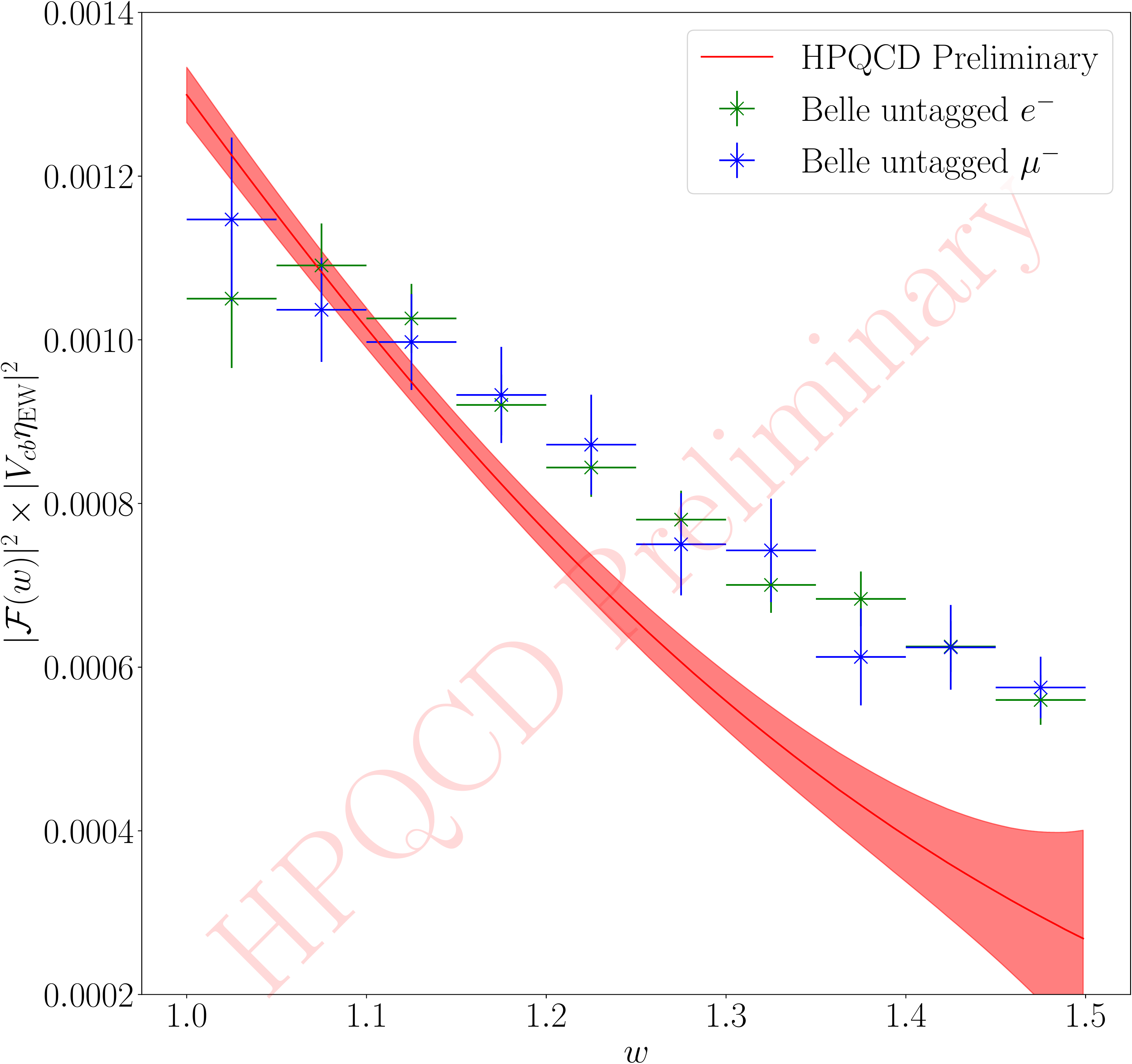}
  \caption{Left: Fermilab-MILC and JLQCD synthetic data points plotted against experimental data. The bands correspond to a fit to a BGL $z$ expansion. JLQCD data does not have an associated band.
           Right: HPQCD result for the decay amplitude, against Belle data. The y-axis scale in both plots is the same to allow for a direct comparison between Fermilab-MILC, JLQCD and HPQCD data.}
  \label{DecayComp}
  \end{center}                                                                                                                                                                  
\end{figure}

When combining the form factors to calculate the decay amplitude, the HPQCD results, like the Fermilab-MILC case, displays a larger slope than experiment, and the resulting curve
consistently plunges below the experimental data (see right pane of Fig.~\ref{DecayComp}).

\begin{figure}[h]
  \begin{center}                                                                                                                                                                  
  \includegraphics[width=0.50\textwidth,angle=0]{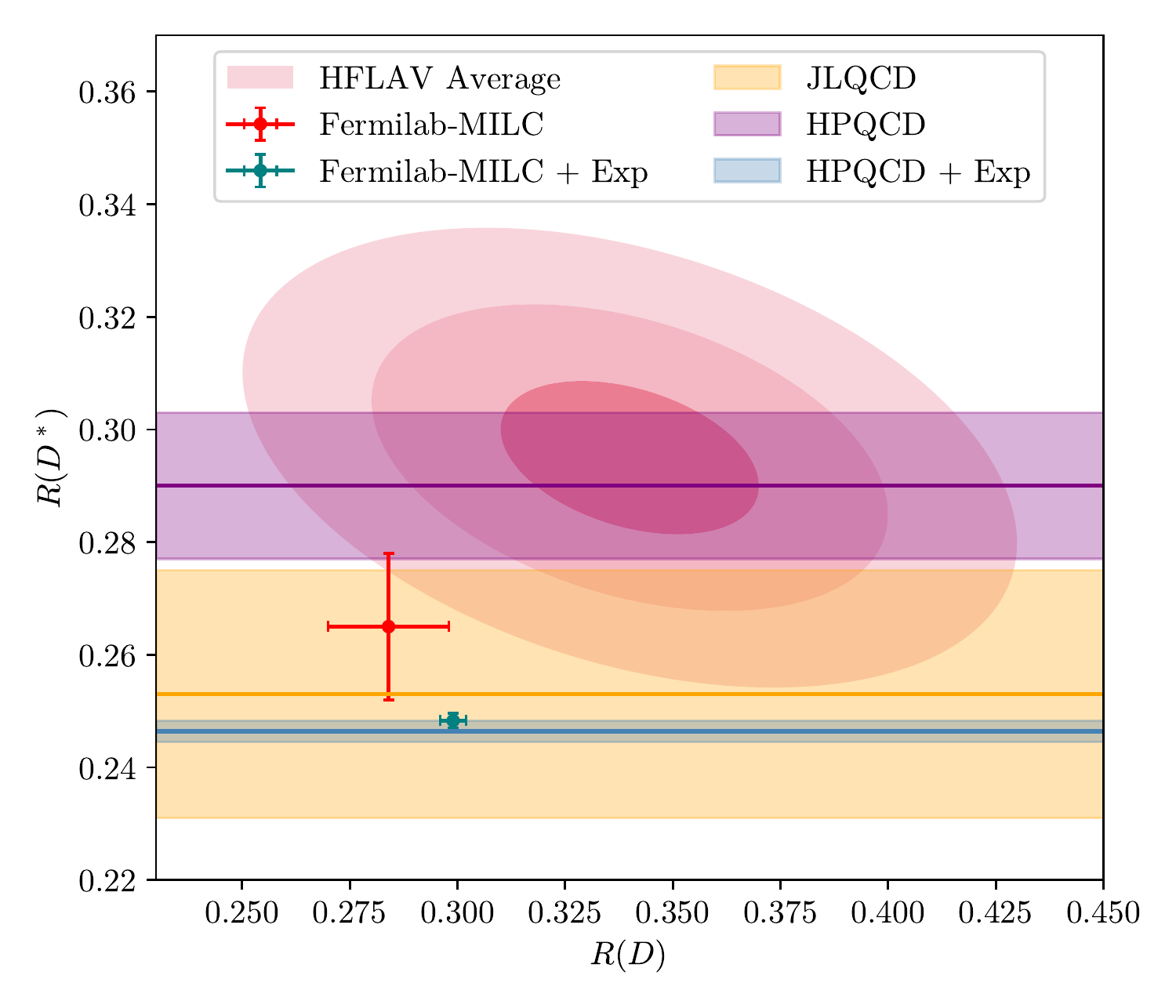}
  \caption{Compilation of results for $R(D)$ and $R(D^\ast)$.}
  \label{RDstComp}
  \end{center}                                                                                                                                                                  
\end{figure}

The collaboration reports a $\chi_{\textrm{aug}}^2/\textrm{dof}\approx 1.3$, where the $\chi^2$ is
augmented, A preliminary joint fit using LQCD and Belle data, and without including the Coulomb factor, yields the preliminary results $|V_{cb}| = 39.2(0.8)\times 10^{-3}$, compatible with
both JLQCD and Fermilab-MILC predictions, and $R(D^\ast)_{\textrm{LQCD}} = 0.280(13)$, $R(D^\ast)_{\textrm{LQCD+Exp}} = 0.2464(19)$, in line with the Fermilab-MILC results. The pure LQCD
determination of $R(D^\ast)$ is perfectly compatible with the current HFLAV average. Results can be checked in Fig.~\ref{RDstComp}.

\section{Future experiments}
\begin{figure}[h]
  \begin{center}                                                                                                                                                                  
  \includegraphics[width=0.60\textwidth,angle=0]{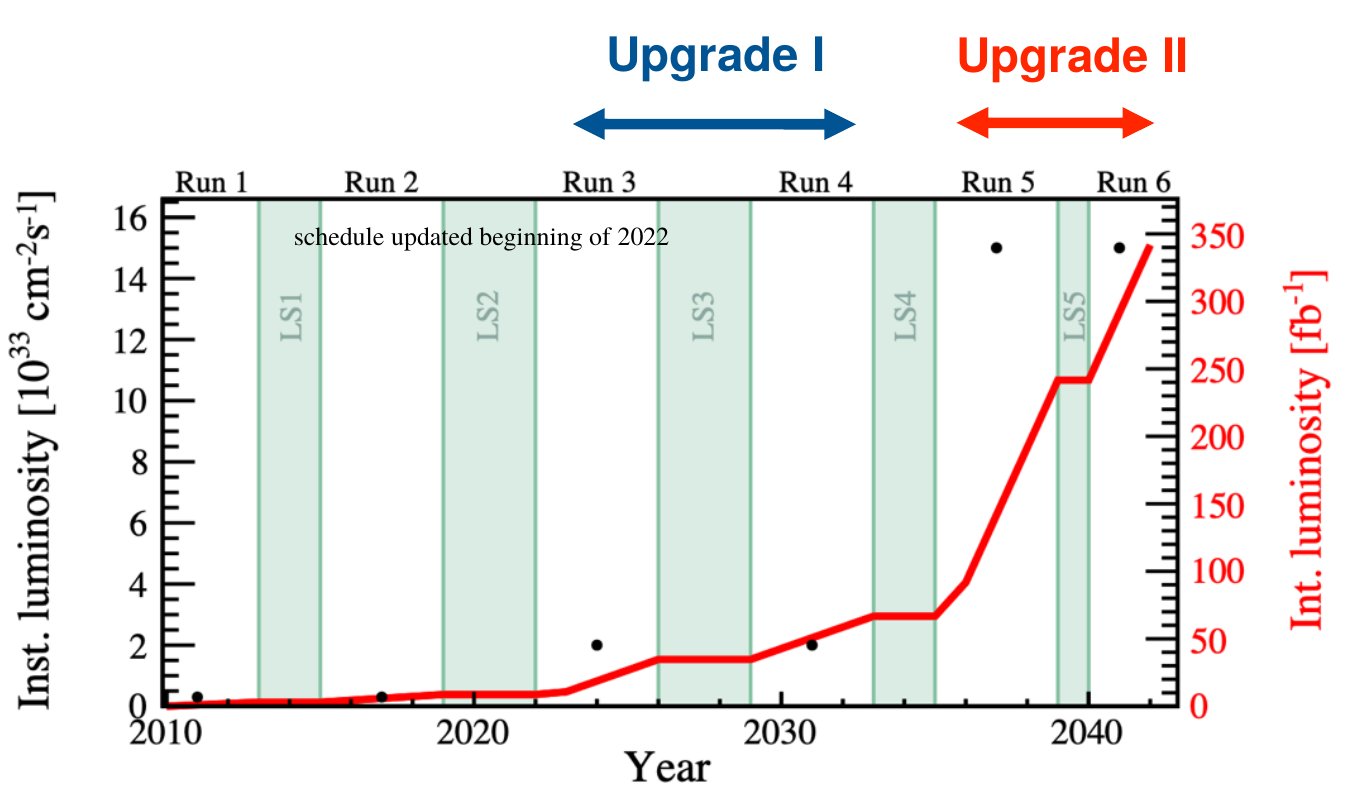} \hfill
  \includegraphics[width=0.35\textwidth,angle=0]{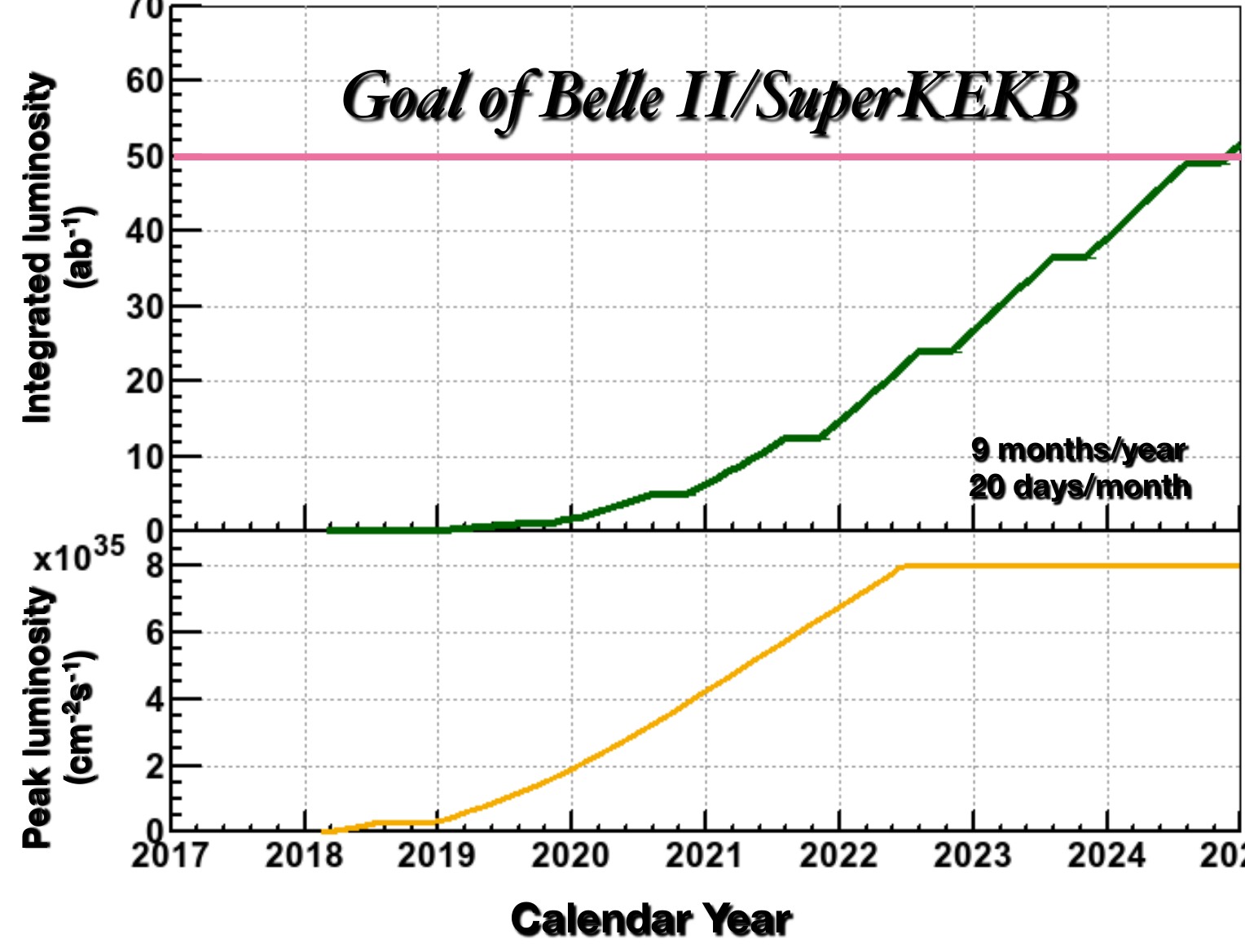}
  \caption{Scheduled integrated luminosity targets for the LHCb and the Belle II experiments. Currently both experiments are running a bit behind schedule due to the pandemic.}
  \label{FPlans}
  \end{center}                                                                                                                                                                  
\end{figure}

Both LHCb and Belle II offer good prospects for the $B\to D\ell\nu$ channels. Both experiments are on track to dramatically improve the precision of existing data by a large factor.
LHCb targets an integrated luminosity of $\approx 350$ fb$^{-1}$ in 2040, whereas Belle is much more ambitious, expecting to reach $50$ ab$^{-1}$ within 5 or 6 years.
The original plans are shown in Fig.~\ref{FPlans}, with the current detectors working behind schedule by one or two years due to the pandemic.
Results have started to come from Belle II. In particular, Ref.~\cite{Belle-II:2022ffa} reports $|V_{cb}|^{\textrm{Untag}}_{B\to D\ell\nu} = 38.3(12)$, and
Ref.~\cite{Moriond:2022sut} gives $|V_{cb}|^{\textrm{Tag}}_{B\to D^\ast\ell\nu} = 37.9(29)$, where the analysis of the $B\to D\ell\nu$ channel is untagged, increasing the statistics and
decreasing the purity, whereas the $B\to D^\ast\ell\nu$ analysis is tagged, with an extremely high purity but low statistics.

\section{Conclusions}
Exciting times are coming in flavor physics. There has been an enormous progress in both theoretical and experimental fronts.
On the theoretical side, the publication of the first complete analysis of the $B\to D^\ast\ell\nu$ form factors in lattice QCD, employing dynamical quarks, is an important milestone
for the community, but two more analyses by the JLQCD and the HPQCD collaborations are close to completion. The existence of several different analyses with different sources of systematic
errors for the same channels allows us to validate results and increases the reliability of the data.
A discrepancy in the shape of the decay amplitude $\mathcal{F}$ between theory (preliminay HPQCD data and Fermilab-MILC data) and experiment is still not well understood, and calls for
further analyses with increased precision. These newer analyses must come soon, as the experiments are steadily increasing their statistics.
On the experimental front the Belle II and LHCb experiments are currently following their scheduled program and promise to deliver high-precision results in the coming years.

These efforts have not been translated into a definite answer to the the very questions that prompted all these developments: we still do not understand why there is a discrepancy between the
inclusive and the exclusive determinations of $|V_{cb}|$, and we need to increase the precisions of both the theoretical and the experimental calculations. The results for $R(D^\ast)$ are also
difficult to read: whereas the pure LQCD values seem compatible with the current HFLAV average, the $R(D^\ast)$ values obtained when combining experimental and LQCD data are in complete tension
with experiment. Certainly more precise analyses are needed.
The increase of precision brings a few new challenges to the table, namely a way to carefully deal with the stability of the $D^\ast$ meson, and the inclusion of QED effects beyond the Coulomb factor.
But looking at current results we expect a small improvement in precision to give hints on the solution to these problems. We can reasonably expect that in the coming years we will see
developments that will help answer the physical questions that have motivated these analyses.

Lattice QCD calculations have also begun to calculate form factors relevant for BSM extensions of the Standard Model, in anticipation of possible solutions for discrepancies found in the future.
Right now, there are not many results available, but the trend is to include these observables on each new calculation, and we should expect high-quality extractions of tensor form factors
for the $B\to D^{(\ast)}\ell\nu$ channels in the near future.

\section*{Acknowledgments}
The author would like to thank T. Kaneko and J. Harrison for the preliminary materials they provided. This work was supported in part by the U.S. National Science Foundation under grants PHY17-19626 and PHY20-13064.

\bibliographystyle{JHEP}
\bibliography{PoSLat22}

\end{document}